\documentclass[12pt,reqno]{amsart}
\usepackage{graphicx,graphics,amsmath}
\usepackage{subcaption}
\usepackage[normalem]{ulem}
\usepackage{t1enc}              %% Ekezetes betuk begepelese:
\usepackage[utf8]{inputenc}    %% unicode
\usepackage{color}
\newcommand{\be}{\begin{equation}}
\newcommand{\ee}{\end{equation}}
\newcommand{\bea}{\begin{eqnarray}}
\newcommand{\eean}{\end{eqnarray*}}
\newcommand{\bean}{\begin{eqnarray*}}
\newcommand{\eea}{\end{eqnarray}}
\newcommand\re[1]{(\ref{#1})}
\def\str{\epsilon} %strain
\def\dst{\sigma} %stress
\def\d{{\rm d}} %differential

\begin{document}
	
	\title[Long term underground measurements in Mátra at MGGL]{First report of long term measurements of the MGGL laboratory in the {M}átra mountain range}
	\author[Barnaf\"oldi et al.]{G.G. Barnaf\"oldi$^1$, T. Bulik$^{7,8}$, M. Cieslar$^{7}$, E.       Dávid$^{1}$, M. Dobróka$^{4}$, \\
	 E. Fenyvesi$^3$, D. Gondek-Rosinska$^9$,  Z. Gráczer$^{2}$, G. Hamar$^1$, G. Huba$^{1}$,\\
	 Á. Kis$^{2}$, R. Kovács$^{1,5A}$, I. Lemperger$^{2}$, P. Lévai$^{1}$, J. Molnár$^3$, D. Nagy$^{3}$,\\ 
	A. Novák$^{2}$, L. Ol\'ah$^{1}$, P. Pázmándi$^{1}$, D. Piri$^{2}$, L. Somlai$^1$, T. Starecki$^{6}$, \\ 
	M. Suchenek$^{6}$, G. Sur\'anyi$^{10}$, S. Szalai$^{2}$, D. Varga$^{1}$, M. Vasúth$^{1}$, P. V\'an$^{1,5A}$, \\
	B. V\'as\'arhelyi$^{5B}$	V. Wesztergom$^{2}$,  Z. Wéber$^{2}$}
	\address{$^1$MTA Wigner Research Centre for Physics, Institute of Particle and Nuclear Physics, 1121 Budapest, Konkoly Thege Miklós út 29-33. \\
	$^2$MTA Research Centre for Astronomy and Earth Sciences, Geodetic and Geophysical Institute, H-9400, Sopron, Csatkai E. u. 6-8.\\
 	$^3$MTA Institute for Nuclear Research, Hungary, 4026 Debrecen, Bem tér 18/c \\
	$^4$University of Miskolc, Hungary, H-3515 Miskolc-Egyetemváros\\
	$^{5A}$Budapest University of Technology and Economics, Department of Energy Engineering, Budapest, Hungary \\
	$^{5B}$Budapest University of Technology and Economics, Department of Engineering Geology and Geotechnics, Budapest, Hungary\\ 
	$^{6}$Institute of Electronic Systems, Warsaw University of Technology, ul. Nowowiejska 15/19, 00-665 Warsaw, Poland\\
	$^{7}$Astronomical Observatory, University of Warsaw, Aleje Ujazdowskie 4, 00478 Warsaw, Poland\\
    $^{8}$Instituto de Astronomía, Universidad Nacional Autónoma de México, Apartado Postal 877, Ensenada, Baja California, 22800 México\\
	$^{9}$Janusz Gil Institute of Astronomy, University of Zielona Gora,  Licealna 9, 65-407 Zielona Góra, Poland,\\
 	$^{10}$MTA-ELTE Geological, Geophysical and Space Science Research Group, Budapest, Hungary}
	\date{\today}
%Keywords: Einstein Telescope, low frequency underground noise, Matra mountain range, long term underground microseismicity, portable muon tomograph

\begin{abstract}
	Matra Gravitational and Geophysical Laboratory (MGGL) has been established near Gyöngyösoroszi, Hungary in 2015, in the cavern system of an unused ore mine. The Laboratory is located at 88~m below the surface, with the aim to measure and analyse the advantages of the underground installation of third generation gravitational wave detectors. Specialized instruments have been installed to measure seismic, infrasound, electromagnetic noise, and the variation of the cosmic muon flux. In the preliminary (RUN-0) test period, March-August 2016, data collection has been accomplished. In this paper we describe the research potential of the MGGL, list the installed equipments and summarize the experimental results of RUN-0. 
	%A novel theoretical framework of noise damping in rock masses is also introduced.
 Here we report RUN-0 data, that prepares systematic and synchronized data collection of the next run period. 
\end{abstract}
\maketitle

\section{Introduction}
% Barnaföldi, G.G., Lévai, P., Vasúth M., Dávid E., Huba G., Kovács R., V\'an P.$^{1}$
%
	
\begin{figure}
\begin{subfigure}{.4\textwidth}
  \centering
 \includegraphics[height=4.5cm]{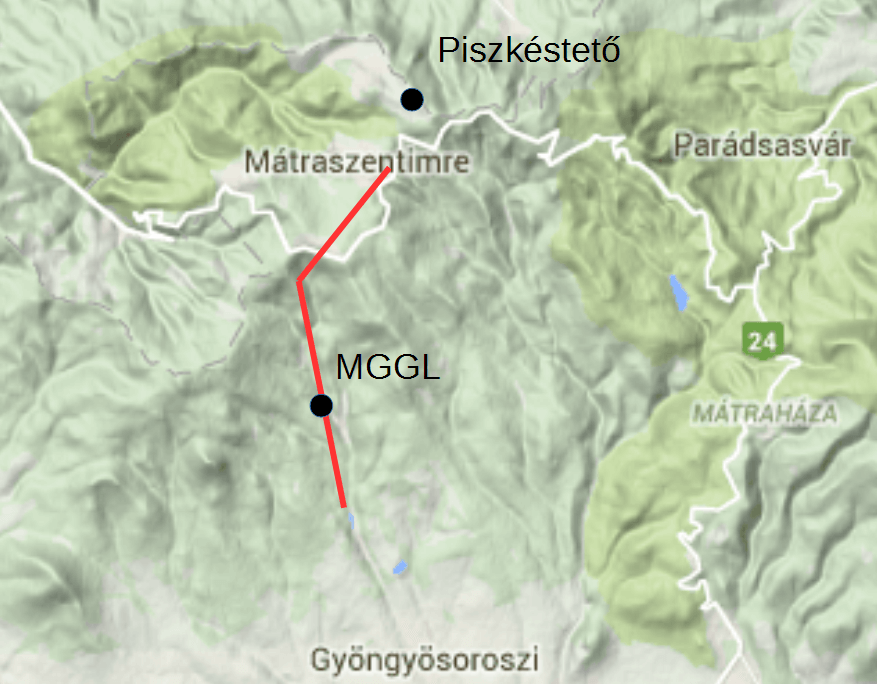} 
 %\caption{}
  %\label{fig:sfig1}
\end{subfigure}%
\begin{subfigure}{.4\textwidth}
  \centering
  \includegraphics[height=4.5cm]{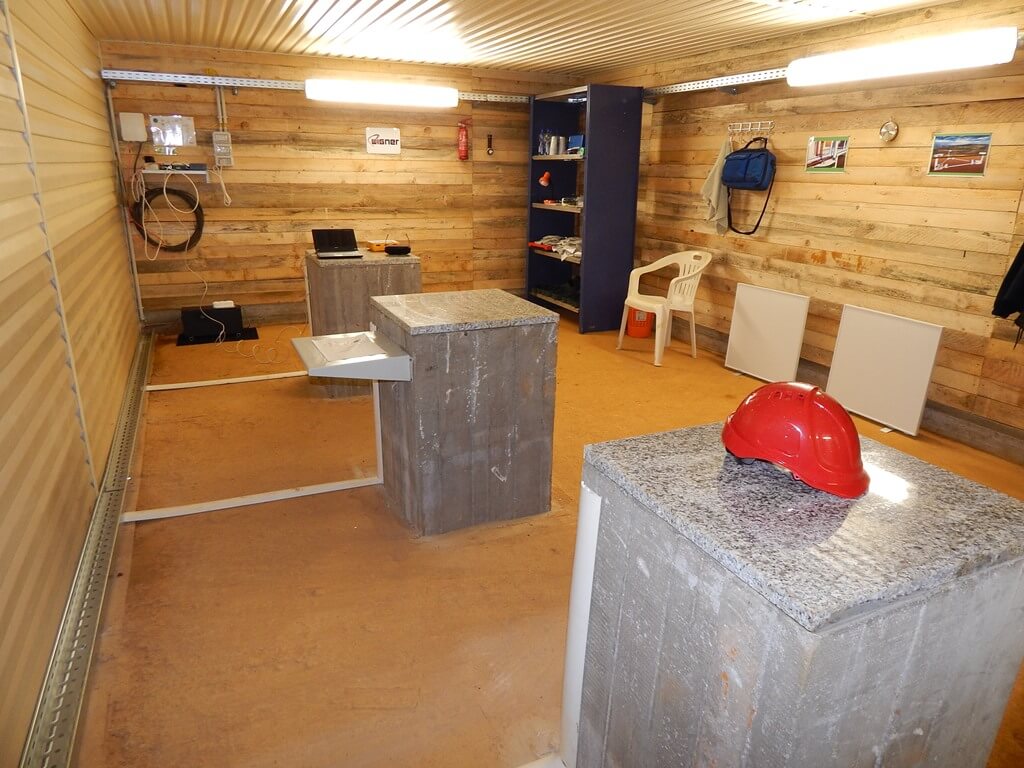}
% \caption{}
 % \label{fig:sfig2}
\end{subfigure}
\caption{Left: The location of Matra Gravitational and Geophysical Laboratory (MGGL) in the  Mátra mountain range and inside the Gyöngyösoroszi mine. The ore mine in Gyöngyösoroszi is indicated by a red line, the black dot displays the 1280~m distance from the entrance. Right: Facilities of seismological measurement inside MGGL.}    
 \label{MGGL_loc}
\end{figure}

The recent discovery of gravitational waves by the LIGO/VIRGO Collaboration \cite{LIGVIR16a,LIGVIR16a2}  generated a focused interest on the further improvements of the detection capability of these ground based facilities, reducing the surrounding environmental noise. A conceptual design study about the feasibility of a third-generation gravitational wave observatory, called the Einstein Telescope (ET), has been completed \cite{ETdes11r}. The underground facility of KAGRA in the Okuhida mountains in Japan is close to perform a test run \cite{Kagra_hp}. As part of the Einstein Telescope design phase, a ground motion study was accomplished in 2010 to determine the seismic noise characteristics at various sites all around the world \cite{Bek13t,BekEta15a}, including a Hungarian site in Gyöngyösoroszi in the Mátra mountain range (Fig. \ref{MGGL_loc}/a). Here the preliminary investigation indicated excellent parameters in noise reduction. 

In 2015 the Matra Gravitational and Geophysical Laboratory (MGGL) of MTA Wigner Research Centre of Physics has been established in the Gyöngyösoroszi ore mine, which is out of operation in the present days.  The Laboratory is located in the coordinates (399 MAMSL,  47$^\circ$ 52' 42.10178", 19$^\circ$ 51' 57.77392" OGPSH 2007 (ETRS89)),  %(399 mBf, 711232.27, 281949.94 EOV), 
along a horizontal tunnel of the mine, 1280~m from the entrance, at 88~m depth from the surface, in the former instruction office near to the first shaft. The Laboratory is near to one of the sites of the above mentioned short term measurements in Refs. \cite{ETdes11r,Bek13t,BekEta15a}. In the MGGL  several measurement platforms were constructed on concrete pillars for long time seismological research (Fig. \ref{MGGL_loc}/b). Furthermore, MGGL has a direct optical data connection to the surface for fast and reliable data transfer. 

The construction of the Laboratory has been finished in February 2016. The first data collection has been started in March 2016 and ended in August 2016. We consider this time period as a test to explore the research potential of the Laboratory and the installed instruments. We name this period as "RUN-0" data collection.  

In this paper we provide a short survey of the geological and rock mass characteristics of the site indicating some further potentially important research lines, and report a preliminary summary of the various measurements in operation. All of them are designed to measure low frequency noises or dedicated to test underground facilities. The participating  equipments are the followings: 

\begin{enumerate}
\item A Guralp CMG-3T seismometer inside the underground MGGL, and a Streickeisen STS/2 seismometer in Piszkéstető, at the surface;
\item A special seismic sensor of the Warsaw University, which was developed specially for testing underground sites for future gravitational wave detector at the critical low frequency range; 
\item An infrasound monitoring system, which was specifically designed to filter false gravitational-wave signals and already applied at aLIGO detectors in the USA; 
\item Lemi-120 extremely low noise and wide frequency induction magnetometers inside the Laboratory and also at a surface site in Piszkéstető;
\item A portable muon telescope (muontomograph), which provides data data to study the density inhomogeneities of the surrounding rock mass. This equipment is designed for underground operation and detects the high-energy cosmic muon flux.
\end{enumerate}

In the following sections we describe the installed facilities in detail and elaborate the RUN-0  data, exploring their nature.

%MGGL, geology, rocks, conditions, maps, equipments and goal. Installed equipments.

\section{Geological background}
%\footnote{Vásárhelyi B.}

The formation of the Mátra is a consequence of several phases of volcanic activity, which lasted for several million years including long quiet intervals. The volcanic activities of the first phase are connected to the beginning of the Tertiary period (Eocene), and their relics can be found on the northern edge of the mountain range.  The main mass of the mountain range was formed in the second phase, during Miocene. The volcano must have been 25 kilometres wide and 2000-2500~m high. The traces of post-volcanic activities in the third phase can be seen on the western and north-western edge of the mountain range. 
Since its evolution, the Mátra has almost constantly been eroding, but in varying degrees; the dominating height of its surface is 700-800 metres which is occasionally interrupted by peaks of 850-1000 metres \cite{Fol88b}.

\begin{figure}[ht]
	\centering
	\includegraphics[width=14cm]{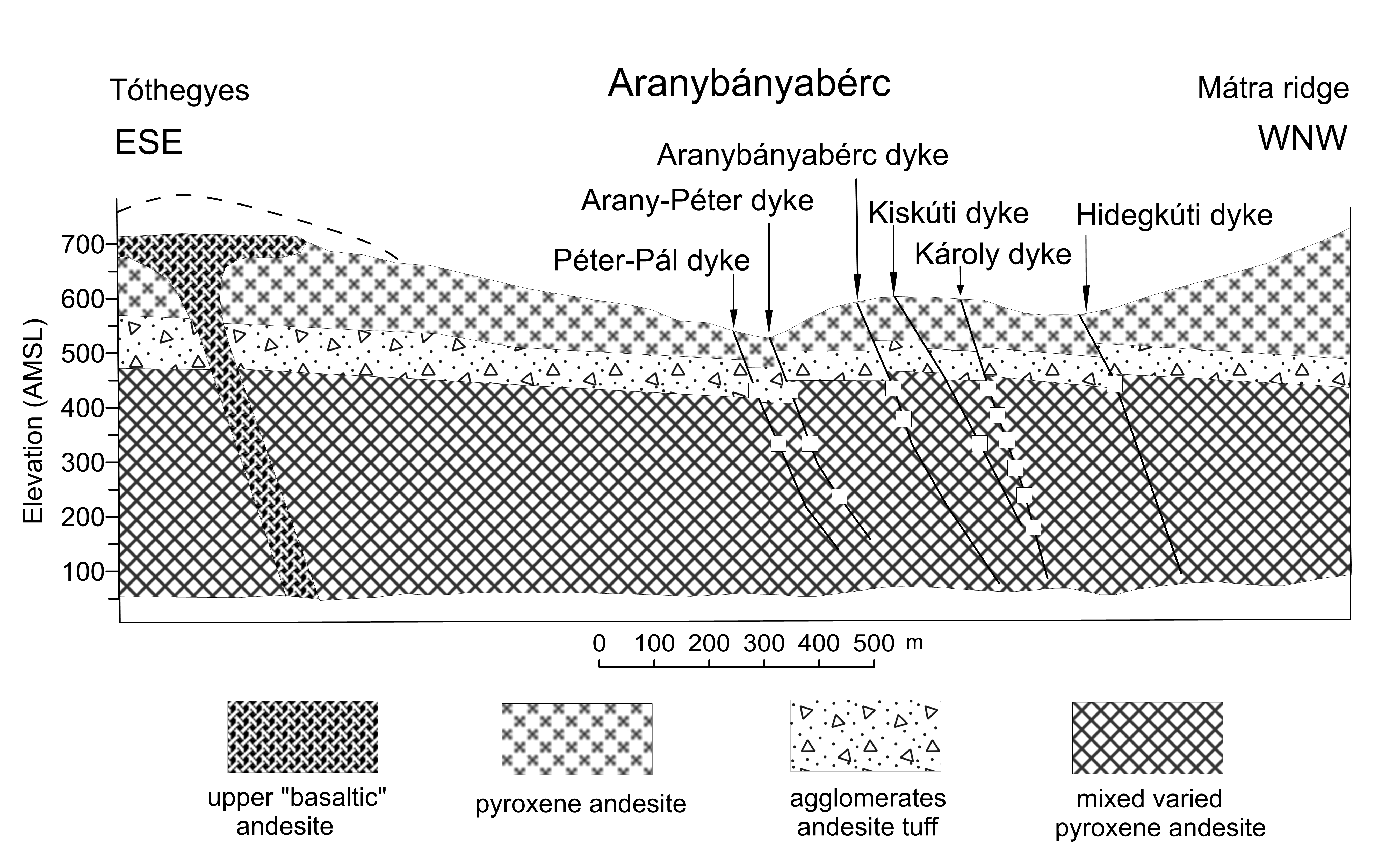}
	\caption{East south-east -- west north-west geological cross-section of the investigated area in the Mátra mountain. The small quadrilaterals show the mine adit. MGGL is located in the Károly dyke, 399~mBf.}
	\label{g1_fig}
\end{figure}

The main rock type  of the Mátra mountains are volcanic rocks, apart from the andesites, include some pro-pyritized andesite (“greenstone”) and also rhyolite. Occasionally we find basaltic tuff and conglomerate. Mainly andesite rock types are in the investigated area (see Figure \ref{g1_fig}).

The Mátra mountain range is along the Darnó-lineament, which is the most important Neogene tectonic component element, trending NNE-SSW cutting the eastern part of the mountains. 

Gyöngyösoroszi is located in North Hungary, at the southern slopes of the Mátra mountains, which is the highest mountain in Hungary. The copper mining activity started as far back as in 1767-1769. At the beginning of 1980, parallel with the mine operation, a special mine inflow operating system was developed against the acidic water. In 2012 the mine was closed and reclamation activity has been started. 

The andesite rock mass is moderately blocky. Around the mine the rock blocks are unweathered (fresh), however, mainly the ruptures and fractures are several times weathered. During the design of infilling the abandoned mine, the rock mass mechanical parameters were calculated by Józsa et al, see Ref. \cite{JozsEta15a}. According to their publication, the following rock mass parameters were used: internal friction angle, $\phi$=25$^\circ$; cohesion, c=1100~kPa; deformation modulus, E=1~GPa; Poisson ratio, $\nu$=0.37; saturated density, $\gamma$=24~kN/m$^3$. 

%\subsection{Discussion}
{ The ET site selection measurements report different noise levels for qualitatively similar rock mass and report that hard rocks are less noisy than softer ones \cite{Bek13t,BekEta15a}. The above data shows that the andesite of Mátra is hard and belongs to the family of less noisy rock types. However, the information above is not sufficient to characterise neither the noisiness nor the noise damping properties of rock masses. The role of the characteristic rock mass properties in noise damping is an open question and may be important in the site selection procedures because of three reasons. 
\begin{itemize}
\item This information can give a preliminary estimation of the optimal depth of the telescope without detailed measurements. 
\item The longest measured characteristic time scale of construction related rock mass relaxation is about a year \cite{KovEta12a1}. The disturbed rock mass relaxes and therefore noisy for a long time after the tunnel construction and this may influence the operation of a gravitational wave detector.
\item The dispersion relation of rocks and rock masses at low frequencies is crucial for the calculation and measurement of gravitational gradient noise filters: the transfer functions cannot be measured by classical seismographic methods. Therefore a combined laboratory and field study is required in each sites. Our more detailed arguments are moved to the appendix.
\end{itemize}
 }

\section{Seismological measurements 1}
%		Wesztergom V., Gráczer Z., Wéber Z., 
% $^2$HAS, Research Centre for Astronomy and Earth Sciences, Geodetic and % Geophysical Institute, \\  
%9400 Sopron, Csatkai E. u. 6-8.

%The Gyöngyösoroszi mine in Hungary is a former lead and zinc mine that is currently being rehabilitated for environmental reasons. It is situated about 70 km north-east of Budapest in the Mátra mountains at an elevation of around 400 m above sea level. The surrounding geology is andesite and andesite tuff. A low-noise broadband seismic station (ET1H) equipped with a Guralp CMG-3T seismometer has been installed along the main drift of the mine in the Mátra Gravitational and Geophysical Laboratory (MGGL). A permanent broadband seismic station (PSZ) of the Hungarian National Seismological Network (HNSN) is situated at the surface, just a few kilometers away from the in-mine station. It is equipped with a high-performance Streckeisen STS-2 seismometer. Both the Guralp and the Streckeisen seismometers have a flat velocity response from ~8~mHz to 50~Hz.

Seismic noise sources are often categorized according to their frequency ranges. For Einstein
Telescope, critical frequency regions are in the domain 0.1 - 10~Hz, because here this is  the primary sensitivity limit together with the Newtonian noise \cite{ETdes11r}. In this region the seismic noise is varying mainly due to microseismic and human activity \cite{BekEta15a} . Noise
sources at frequencies below 1~Hz are dominantly natural, depending on oceanic
and large-scale meteorological conditions. Around 1~Hz wind effects and local
meteorological conditions show up, while for frequencies above 1 Hz additional
sources (besides natural) are related mainly to human activities. Such noise
is termed ‘cultural noise’ or ‘anthropogenic noise’. It is therefore important to choose a site location far from oceans, as well as present and future human activities.

Peterson \cite{peterson1993} catalogued noise power spectral density (PSD) plots from 75 seismic stations distributed worldwide. Several years of data were collected
(about 12,000 spectra in total). From the upper and lower bound of the combined
data of the selected stations, Peterson derived the New High/Low Noise Model
(NHNM/NLNM).

Microseismic ground motion is a prominent feature for frequencies around 0.07 Hz
and 0.2 Hz. The small lower-frequency peak correlates with the frequency of
coastal waves, where the ocean wave energy is converted into seismic energy
through either vertical pressure variations or from the surf crashing onto the
shore. The larger peak at about 0.2~Hz originates from standing ocean waves that
couple to the continental shelf. The standing waves are generated by
superposition of ocean waves of equal period traveling in opposite directions.

The large interferometric detectors GEO600, LIGO and Virgo are placed
on the surface of the earth and, consequently, are sensitive to seismic
disturbances. From ground motion measurements it is clear that the level of
seismic noise in underground environments is usually much less than the one the
surface. Thus, the need to reduce the seismic noise to achieve the desired
sensitivity for the Einstein Telescope seems to be fulfilled by choosing an
underground site where the low seismic activity and the uniformity of the
underlying structure may play a dominant role in the site selection
process.

According to the above short discussion, the aims of our seismological
measurements at the MGGL are seismic noise characterization of the mine and the
study of noise attenuation with depth.  In order to achieve the above goals,
seismic noise measurements are carried out using both the surface station and the in-mine station. Data obtained from these two stations are then compared with each other.

%Az elejéről áthozva.
The low-noise broadband seismic station (ET1H) is equipped with a Guralp CMG-3T seismometer,  installed along the main drift of the mine in the Mátra Gravitational and Geophysical Laboratory. A permanent broadband seismic
station (PSZ) of the Hungarian National Seismological Network (HNSN) is situated
at the surface, just a few kilometers away from the in-mine station. It is
equipped with a high-performance Streckeisen STS-2 seismometer. Both the Guralp
and the Streckeisen seismometers have a flat velocity response from ~0.008~Hz to 50~Hz.

\begin{figure}[ht]
	\centering
	\includegraphics[width=14cm]{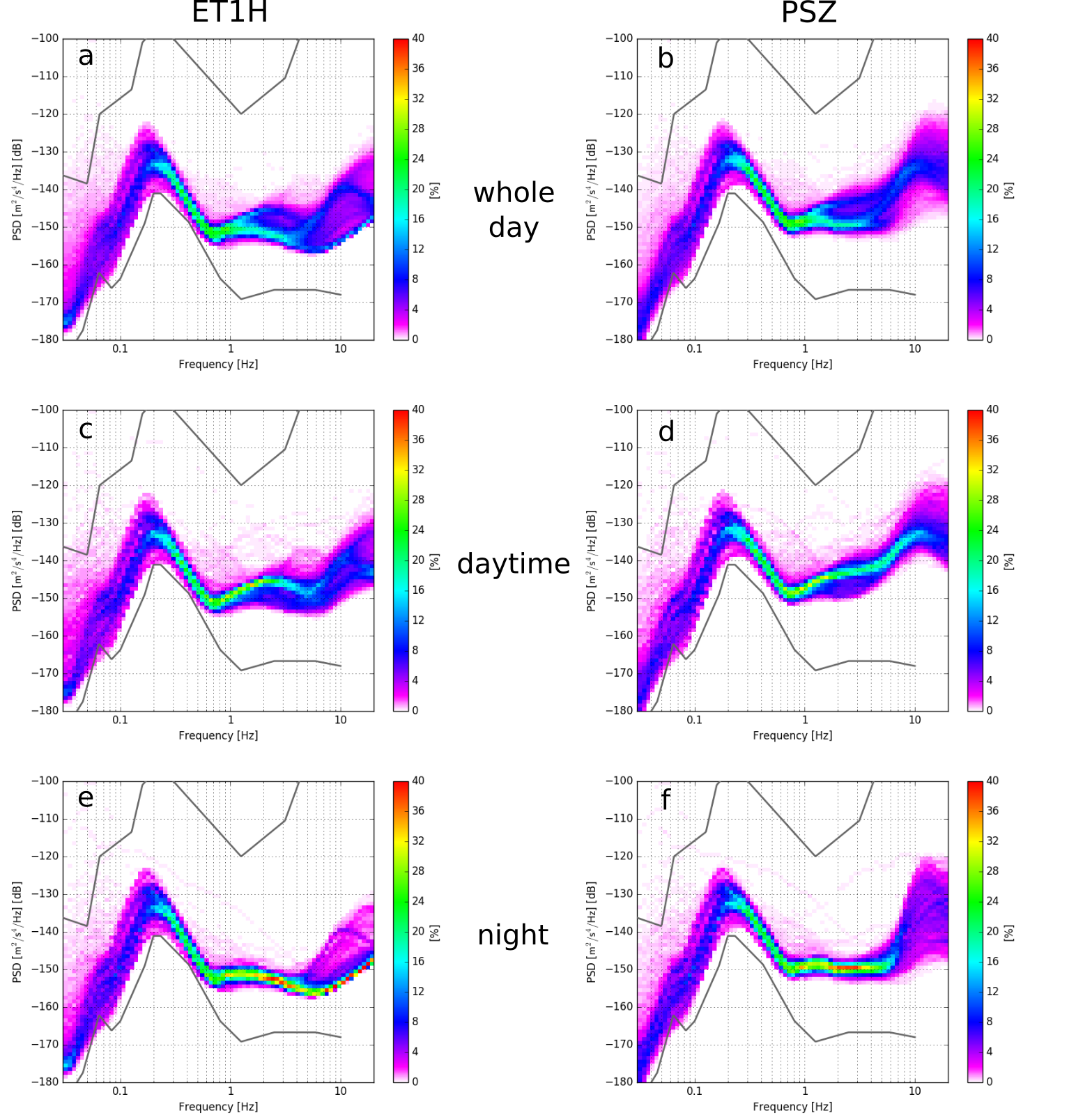}
	\caption{
Vertical component power spectral density (PSD) histograms for the underground
site ET1H (left column) and the surface station PSZ (right column) of the RUN-0 seismometer data. The
overall PSDs (first row), the daytime PSDs (second row), and the night PSDs
(third row) are illustrated separately. Thick grey lines represent the low/high
noise models of Ref. \cite{peterson1993}. Colors indicate histogram levels.}
	\label{seismo1}
\end{figure}

\begin{figure}[ht]
	\centering
	\includegraphics[width=14cm]{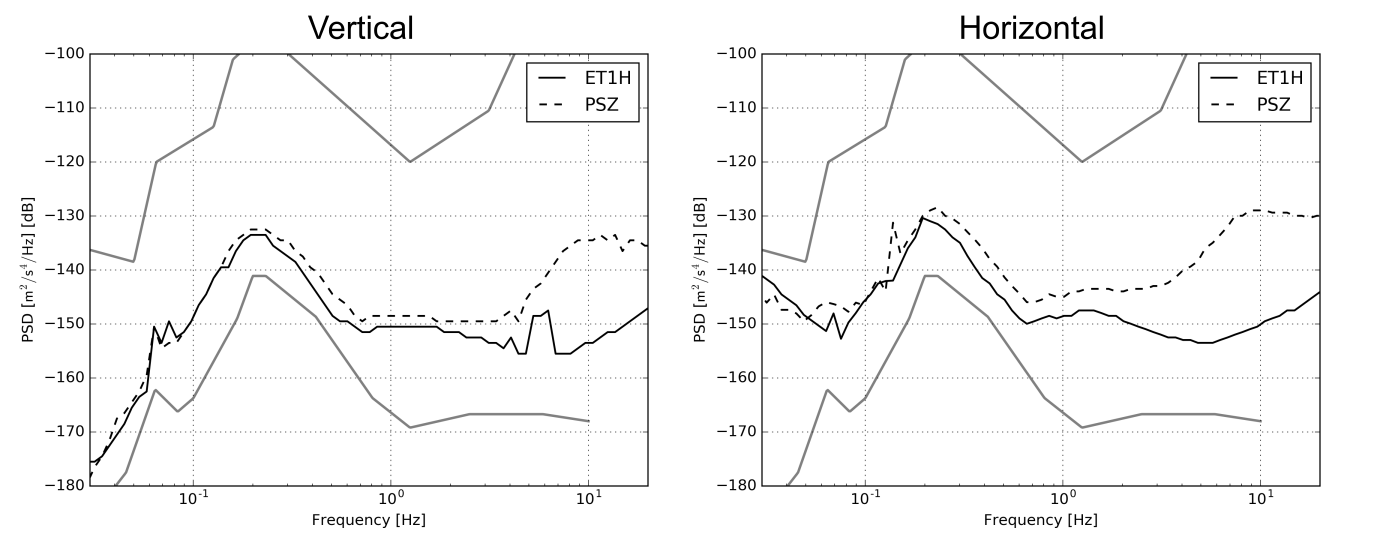}
	\caption{
Comparison of the vertical (left) and horizontal (right) PSDs at the underground
site ET1H (solid line) and surface station PSZ (dashed line) RUN-0 data. Lines represent
the mode of the corresponding PSD histograms after three-point moving average
smoothing.}
	\label{seismo2}
\end{figure}

\begin{figure}[ht]
	\centering
	\includegraphics[width=14cm]{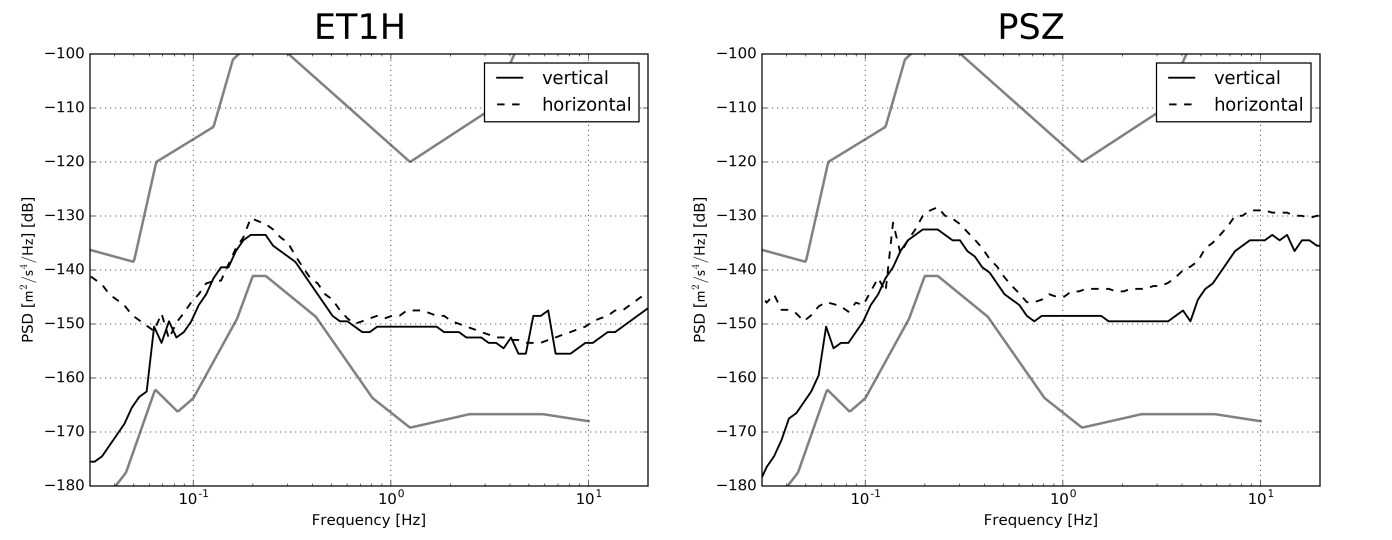}
	\caption{
Comparison of the vertical (solid line) and horizontal (dashed line) PSDs at the
underground site ET1H (left) and the surface station PSZ (right) of RUN-0 data. Lines
represent the mode of the corresponding PSD histograms after three-point moving
average smoothing.}
	\label{seismo3}
\end{figure}

Measurements were obtained over a three-month period (1\textsuperscript{st} March - 31\textsuperscript{st} May 2016). In all of the results presented below, fast Fourier transform (FFT) was performed to obtain power spectral densities (PSDs) in $\mathrm{(m^2/s^4)/Hz}$. For comparison, the high- and low-noise models are also plotted in Figs \ref{seismo1}-\ref{seismo3}. Spectral variation plots are used here to not only show the amplitude of the seismic signal, but also present here how much time, as a percentage, is spent at a certain level. This is indicated by the color of the plot (see sidebar).

The PSD plots obtained for the vertical component waveforms at the surface station PSZ and the underground site ET1H are summarized in Fig.~\ref{seismo1}. the first wo compare the overall spectral variation at the two sites, whereas the figure pairs at the middle depict the PSDs in daytime (between 8 and 16 hours local time) and the Figures below at night (between 22 and 6 hours local time). The Figures clearly show that the noise level is usually lower for the in-mine station than for the surface site, especially above 0.7 Hz. In this frequency range, the effect of noise is more pronounced on the surface than in depth. Above 1 Hz, where seismic noise is mostly due to human activity, the noise level and its scatter are higher in daytime than at night.

In order to better emphasize the difference between the surface and the underground sites, Fig.~\ref{seismo2} shows the mode of the PSD histograms at the two stations for the vertical and horizontal components. Below the microseismic peak at 0.2~Hz the difference between the two stations is negligible. Between 0.2 and 5~Hz, however, the vertical component noise level at ET1H is about 2-5~dB lower than that at PSZ, whereas at higher frequencies the difference increases to as large as 20~dB. Regarding the horizontal component, the difference between the two stations is more pronounced: between 0.2 and 2~Hz it reaches $\sim$5~dB, in the frequency range of 2-4~Hz it is as large as $\sim$10~dB, and at higher frequencies it exceeds 20~dB. Accordingly, cultural noise generated by human activities has mainly horizontal component.

Figure~\ref{seismo3} compares the vertical and horizontal component noise levels for the two seismic stations. At the underground site ET1H, above the lower microseismic peak at 0.07~Hz, the vertical and horizontal noise components are practically the same. At the surface station PSZ, however, the noise level of the vertical component is about 5-10~dB lower than that of the horizontal components together  for frequencies above 1~Hz. This observation also proves that at the surface human activities mostly generate horizontal-component noise. Both components are important for the ET design.

\section{Seismological measurements 2}
%Please indicate the authors of the section together with their affiliation, like it is  in the introduction. 
%Suchenek, M.$^{1}$, Bulik, T.^{2,3} Rosinska D.$^4$, T. Starecki$^{1}$, M. Cieslar$^{2}$
%
% $^{1}$ Institute of Electronic Systems, Warsaw University of Technology, ul. Nowowiejska 15/19, 00-665 Warsaw, Poland
% $^{2}$ Astronomical Observatory, University of Warsaw, Aleje Ujazdowskie 4, 00478 Warsaw, Poland
% $^{3}$ Instituto de Astronomía, Universidad Nacional Autónoma de México, Apartado Postal 877, Ensenada, Baja California, 22800 México
% $^{4}$ Janusz Gil Institute of Astronomy, University of Zielona Gora,  Licealna 9, 65-407 Zielona Góra, Poland
%

\begin{figure}
\includegraphics[width=0.45\textwidth]{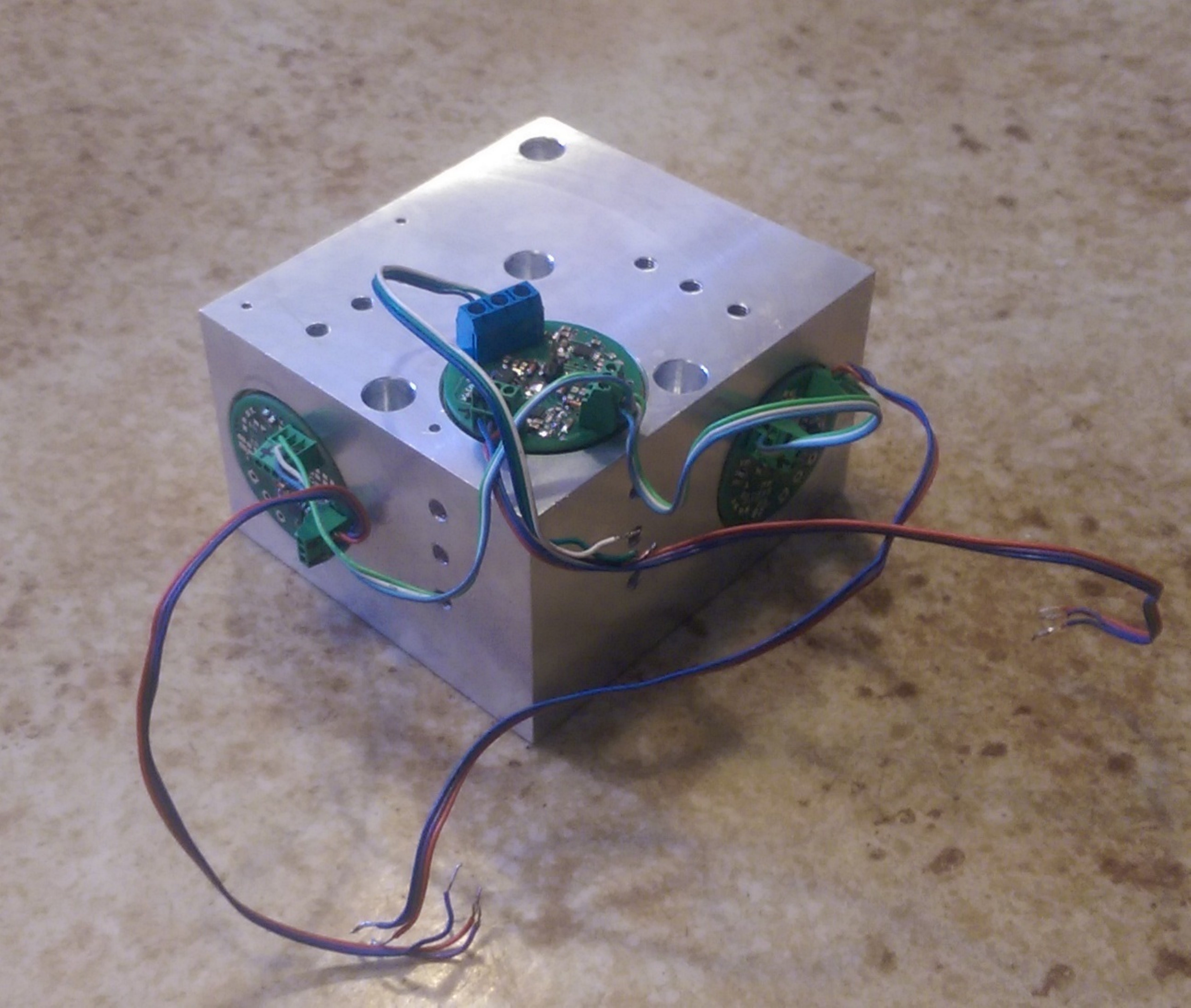}
\includegraphics[width=0.45\textwidth]{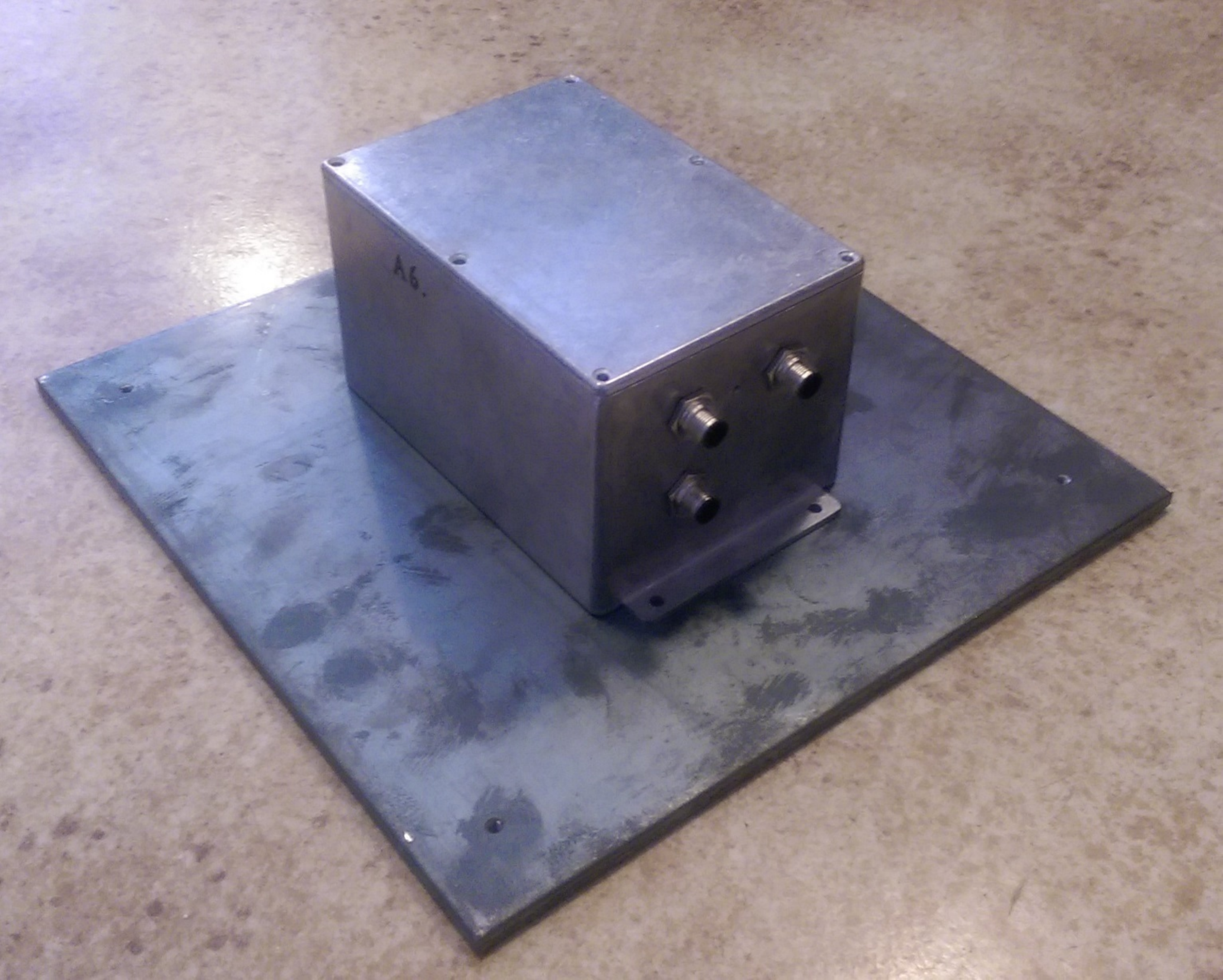}
\caption{The left panel shows the three geophones in the metal housing, while the right panel presents the entire detector mounted on a metal plate.}
\label{Sensor}
\end{figure}
\begin{figure}
\includegraphics[width=0.8\textwidth]{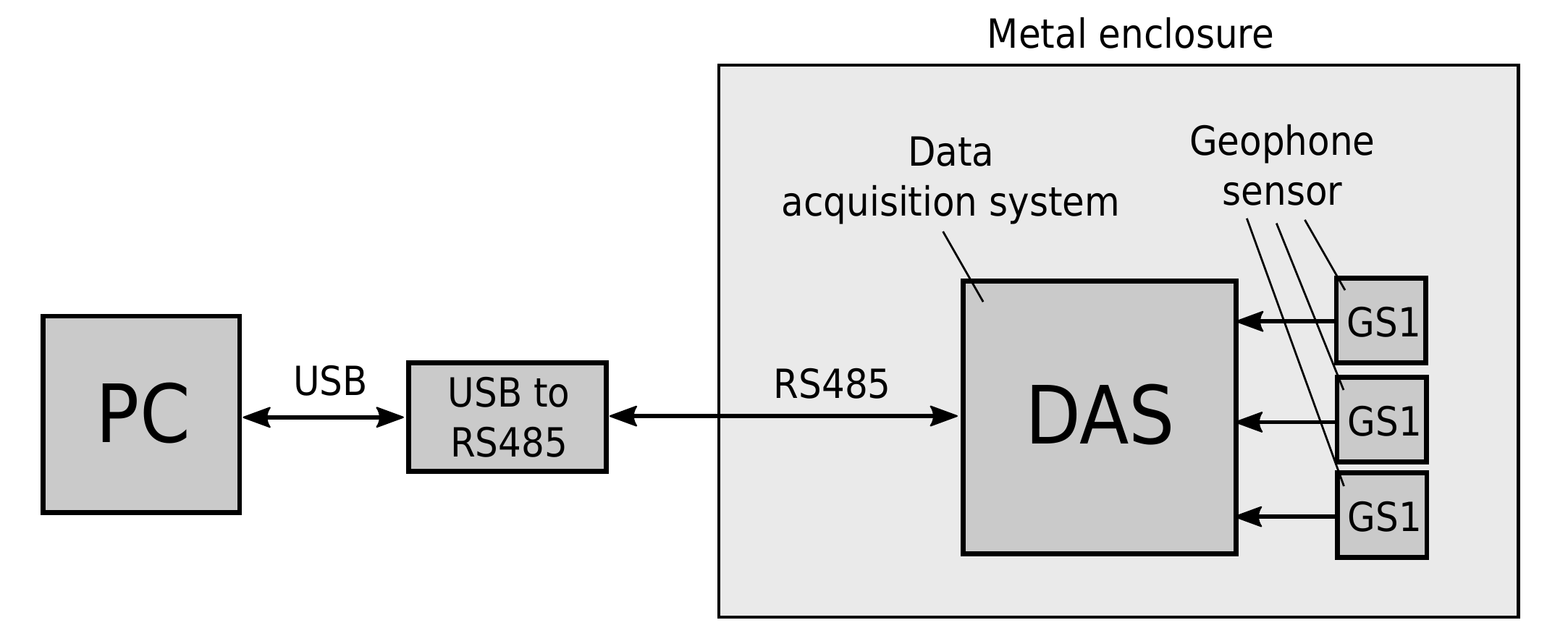}
\caption{The block diagram of the data acquisition system.}
\label{sensor-block}
\end{figure}

A second, independent seismic noise data acquisition was performed with a custom made seismic sensor of the Warsaw University. The measurement system uses three-axis seismometer, one for vertical and two for horizontal measurements. We are using the LGT-2.5 and LGT-2.5H  geophones  with the resonance frequency of
$ 2.5$~Hz. The three sensors are  mounted in metal housing, shown  in Fig. \ref{Sensor}. The analogue signal from the geophones is connected to the data acquisition system placed inside the metal housing, see the right panel in Figure \ref{Sensor}. The block diagram of the seismic measurement system is presented in Figure \ref{sensor-block}.

%have sensitivity 2 V/cm/s

%%The data acquisition system is responsible for periodic sampling of the signal from sensors. The frequency can be set in range from 125 Hz to 1 kHz. The analogue-to-digital converter used in the system  allows to measure the signal in the range of $\pm 2.5$V with the resolution of 32 bits. The data acquisition system records the data and sends it in blocks to the computer via an RS485 interface. The signal can then be received by any standard PC with USB interface with the use of a “USB-RS485” converter. The data are sent to the computer every 28 seconds. The same data are recorded on a  SD card placed in the seismometer as well. The maximum capacity of the card supported by the system is 32 GB which allows  for approximately  eight months of constant recording. The block diagram of the seismic measurement system is presented in Figure \ref{sensor-block}. The data acquisition  system can be extended and the measured seismic signal can be synchronized between several  seismometers via the  RS485 interface. In such configuration additional synchronization connector is used in each seismometer. 

The instrument was calibrated by comparing the reading with a Trillium broad band seismometer, and its sensitivity was assessed by measurements of the noise in a seismically isolated vacuum chamber. The sensitivity  at 1~Hz was found to be $\approx 2\times 10^{-10}$m\,s$^{-1}$\,Hz$^{-1/2}$. 

\begin{figure}
\includegraphics[width=0.45\textwidth]{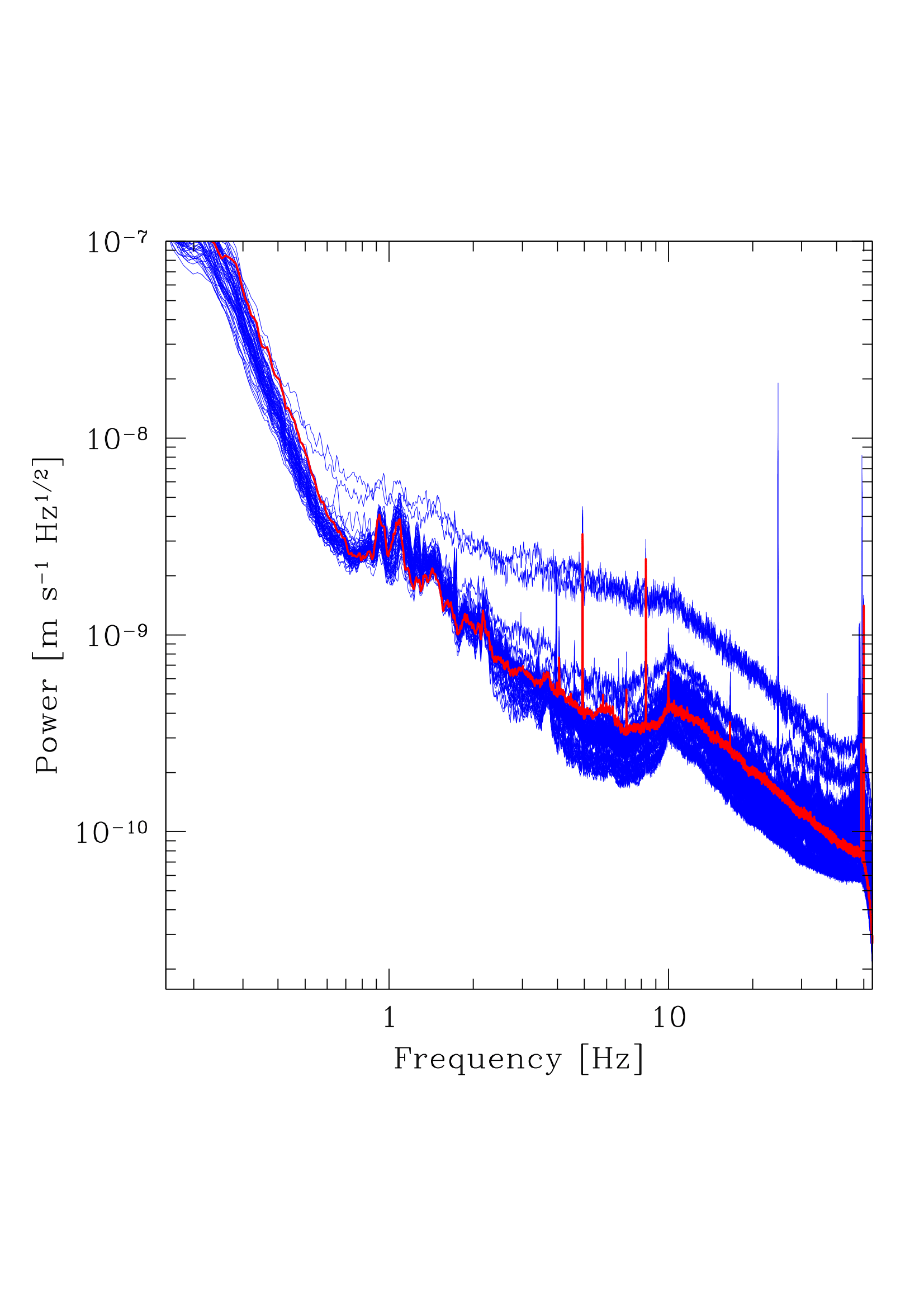}
\includegraphics[width=0.45\textwidth]{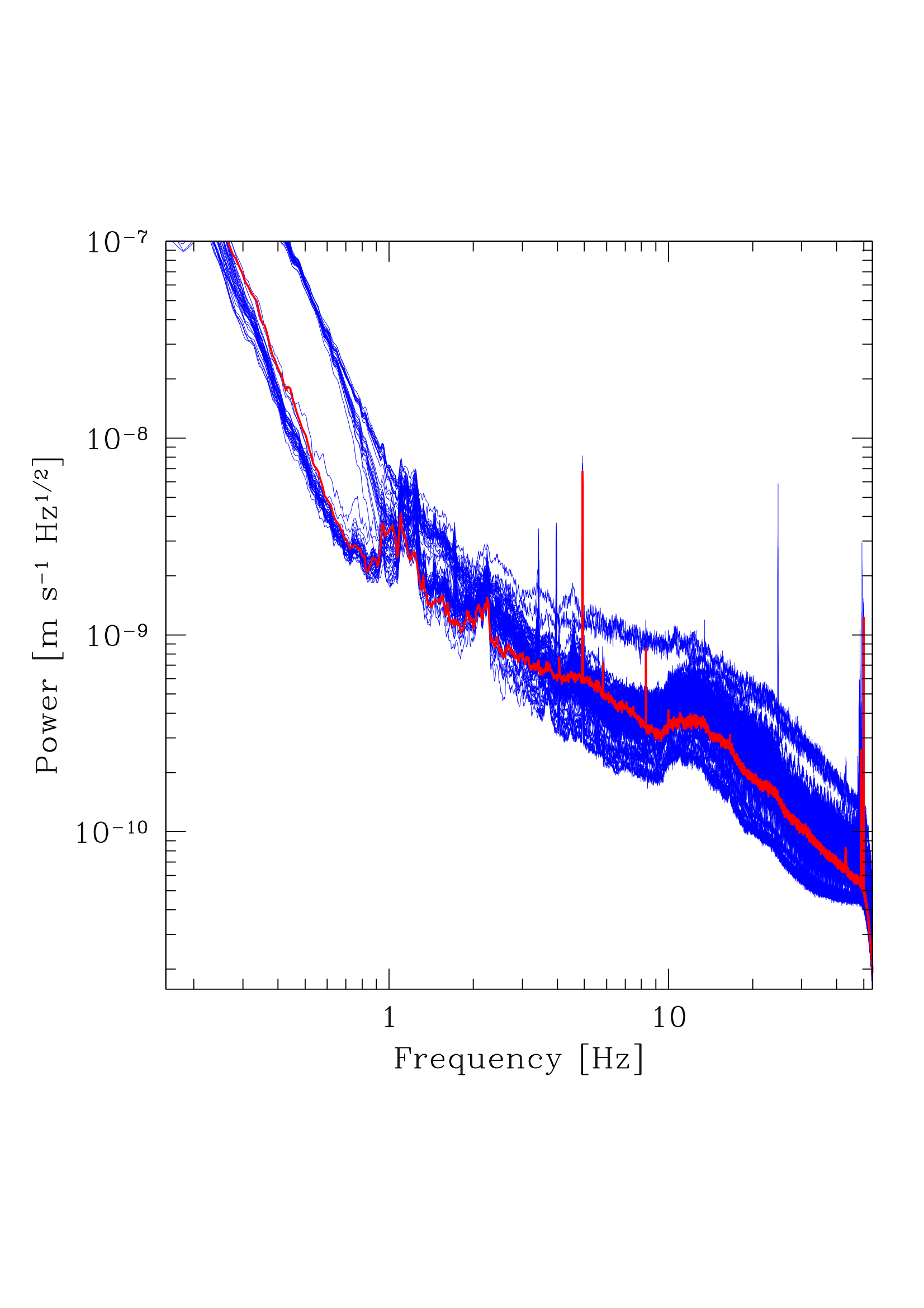}
\caption{The daily power spectra of velocities of the RUN-0  data of seismic sensor of the Warsaw University. The left panel shows the velocity amplitude spectra in the horizontal direction while the right panel present the vertical ones.  The data of the quiet day - 16\textsuperscript{th} July  -  is shown in red. }
\label{figwaw}
\end{figure}

We have started gathering the data on May 24\textsuperscript{th}, 2016. In this paper we present the early analysis of the first 77 full days of data, starting on 25\textsuperscript{th} May, 2016, until 8\textsuperscript{th} August, 2016.  The instrument performed flawlessly during this period. For the purpose of data analysis the data of each day were divide into the 685 chunks with the length of 16384 samples. We have calculated the velocity amplitude spectra of each chunk, and then calculated the daily average velocity amplitude spectra. We present the results in Figure \ref{figwaw}. The level of the seismic noise is approximately $2-3\times 10^{-9}$m\,s$^{-1}$Hz$^{-1/2}$. There were two days  with increased noise level in the 1 to 10Hz band. Apart from that the seismic noise level is very stable, and the day-to-day variation of the noise is smaller than  a factor of two.

We find several lines in the data. There is the omnipresent 50Hz line with a 25Hz sub-harmonic. However, there are also quite narrow lines at 4 and 5 Hz in the  spectra.  The identification of the lines is not clear but we believe that they are caused by anthropogenic noise. One hint of the origin of lines comes from analysing the 16\textsuperscript{th}  July  data. This has been agreed to be a clear day when all non essential machinery is turned off in the mine. The data of 16\textsuperscript{th} July is presented as red lines in Figure \ref{figwaw}. Some of the lines are not present in the averaged spectrum of that day, which clearly supports the interpretation of their origin in machinery operating in the mine. Additionally it is worth noting that the overall level of the noise on the quiet day does not differ strongly from other spectra. This means that the artificial seismic noise in the mine is constrained to narrow bands. A detailed analysis of the data will presented in a forthcoming paper.

\section{Infrasound monitoring}
%Molnár J., Fenyvesi E., Nagy D. $^{2}$ \\
%$^2$HAS, Institute for Nuclear Research \\  
%Hungary, 4026 Debrecen, Bem tér 18/c

Beside seismic motion, the pressure waves in the air make vibrate the components of interferometrical gravitational-wave detectors, especially the outer suspension points of the in-vacuum system and other components that are not protected by vacuum \cite{Effler}.  While these soundwaves increase noise, they also can generate false positive gravitational-wave (GW) signals, which have to be distinguished from real signals \cite{LIGVIR16a1}.

The measuring range of current GW detectors (aLIGO,	AdVirgo, KAGRA) is 10~Hz-10~kHz \cite{MarEta16m}.
At aLIGO the microphones, which are sensitive in the critical low frequency range   10~Hz-30~Hz are able to detect sound waves which cause false positive GW signals. After detecting these false signals, special algorithms can veto them. Since Einstein Telescope (ET) is designed to detect GW signals in the frequency range 1~Hz-30~Hz, monitoring the pressure waves of air that is lower in frequency than 20~Hz (infrasound) will be essential in order to use veto algorithms in this range.

Beside these effects, infrasound also contributes to gravity gradient noise of the detectors. This type of noise is caused by moving masses in the ground and air resulting fluctuating Newtonian gravitational forces around the interferometers which can modify the positions of test masses. Though, this effect is intensively examined theoretically \cite{Creighton,Driggers}, new measurements with suitable sensors are needed to investigate the effects of infrasound on GW detectors. 

\subsection{Infrasound monitoring system}
We have developed a new infrasound monitoring system at ATOMKI, Debrecen for the above purpose. The system consists of a capacitor infrasound microphone and a data acquisition system. Since the pressure resolution of the microphone is on the order of 1~mPa within the range of 10~mHz - 10~Hz, it is suitable to measure air pressure fluctuations in the physical environment of ground-based and underground interferometric GW detectors.

The design is based on the design concept expounded in Ref. \cite{LosAlamos}. The microphone is sensitive to the changes of air pressure with a flexible diaphragm, which separates a reference volume from the outside environment. The reference volume is connected to the environment through a small capillary.  The diaphragm moves in response to the difference between the pressure of the environment and the reference pressure.  These vibrations are converted to an analog electronic signal by a sensor, and sampled to produce time series of amplitudes linearly proportional to the infrasound wave amplitude. The sampling frequency is 1024~Hz.

%%A 16-Bit ADC and a microcontroller convert the analog signal to digital one, according to the RS-485 standard by using universal asynchronous receiver/transmitter (UART). Then the RS-485 signal is transmitted to a digital to analog converter (DAC), which produces an analog signal and transmits the RS-485 signal, too. Then the RS-485 signal is converted to USB. USB data is collected to an SD card by a software running on a Raspberry Pi computer~\cite{raspberry}.  The software assigns time stamps to the the data, too. Exact timestamps are ensured by a Time Protocoll (NTP) client running on Raspberry Pi. Data of one day length is stored in a binary file on the SD card, and can be downloaded via internet connection.

\subsection{Data processing and results}
The infrasound monitoring system was installed at MGGL in order to characterize the infrasound background at the Laboratory and to test the capabilities of the measuring system. Data collected between 16\textsuperscript{th} June and 21\textsuperscript{th} August 2016 were processed. 
%%Binary data of each day were downloaded from the SD card and then converted to comma separated (CSV) file. During file conversion, the raw data was downsampled to 256~Hz. 
In the mine reclamation work was done during daytime, so only the first seven hours of each day was used to compute representative Pressure Amplitude Spectral Density (PASD).  PASD plots depict the power of the infrasound signal of a given frequency. 
%%The units of PASD are in Pa Hz$^{-1/2}$.

\begin{figure}[ht]
	\centering
	\includegraphics[width=13cm]{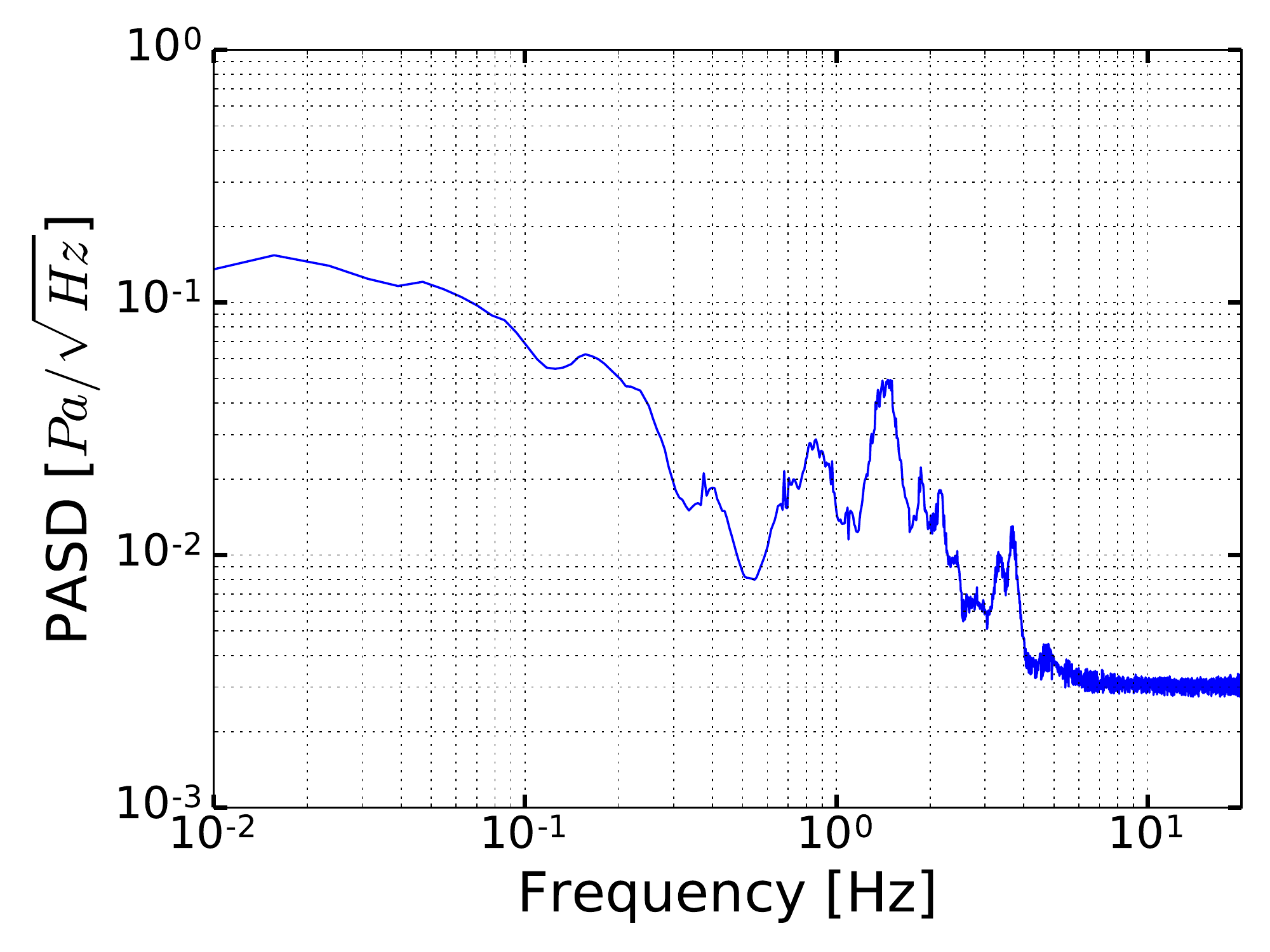}
	\caption{Representative pressure amplitude spectral density (PASD) of low-frequency pressure waves registered by the infrasound monitoring system in RUN-0 data collection of MGGL. The solid curve represent the mode, or most common PASD value corresponding to a given frequency}.
	\label{ASD} 
\end{figure}

Data processing method disclosed in Ref. \cite{BekEta15a} was implemented in a Matlab software. 
%% The resulting average PASDs corresponding to separate time segments of 1792~s length was stored in CSV files. 
Then, the most common PASD value (mode) related to a given frequency was computed (see Fig. \ref{ASD}). Studies aimed to separate noisy and quiet periods and the identification of infrasound noise sources is under progress. The peaks of the plot may be due to tunnel resonances. In order to examine this possibility, more targeted investigations are planned with several measuring sites along the tunnels, with an advanced infrasound monitoring system, including more microphones. %The plot shows that infrasound noise is significantly mitigated by the rock above MGGL, since the PASD values measured at ground-based GW detectors can not be decreased below $10^{-4}$~Pa/$\sqrt{Hz}$ at any frequency interval \cite{Fiorucci}. In MGGL, ASD remains below $10^{-4}$~Pa/$\sqrt{Hz}$ at frequencies higher than 2~Hz, which is the lower measuring limit of ET, and even can be decreased below $10^{-5}$~Pa/$\sqrt{Hz}$.
It can be assumed, that the appropriate choice of the geometry of the tunnels for ET can result  lower values, since laboratory specific resonances of MGGL can be avoided.

\section{Electromagnetic measurements}

% Wesztergom V., Lemperger I., Novák A., Piri D., Kis Árpád, Szalai S.
% HAS, Research Centre for Astronomy and Earth Sciences, Geodetic and Geophysical Institute
% H-9400, Sopron, Csatkai E. u. 6-8.

The comprehensive geophysical study of the site involves detailed electromagnetic investigation of the local environment too. On the one hand the so called magnetotelluric exploration has been proposed to recover the spatial distribution of the electric conductivity in the close vicinity of the subsurface laboratory, MGGL. On the other hand an electromagnetic noise test has been performed focussing to the upper ULF (Ultra Low Frequency)-lower ELF (Extremely Low Frequency) range. The following subsection provides a brief description and specification of the hardware instruments and software environment applied in the course of the electromagnetic noise measurement campaign. 

%\subsection{Electromagnetic noise test}
The frequency range of interest in respect of the Einstein Telescope is the lower part of the ELF range, with critical focus on 1-10~Hz window. Detailed understanding and modelling of the propagation characteristics of natural and man-made electromagnetic (EM) signals in the subsurface is also of great importance to ensure the expected sensitivity of the proposed gravitational wave detector.
The sources of the natural signal in the ELF range are mostly related to the global thunderstorm activity and local individual lightning discharges. The former results in the so called Schumann background, which is characteristic issue of the electromagnetic excitation of the Earth-ionosphere cavity. The latter occurs occasionally, results in few millisecond long broadband transient impulse and related to local meteorological conditions.
The investigation of the EM attenuation characteristics of the overlying rock in the frequency range of interest is feasible by long turn multisite monitoring of the horizontal magnetic components of the natural source ELF signals. In the frame of the project the empirical EM transfer function of the overlying andesite block is proposed to be reconstructed by means of comparative spectral analysis of the EM signal recorded at the surface and at the subsurface location.
The electromagnetic noise test proposed in the MGGL consists of parallel observation by means of a underground (MGGL) and a surface (Piszkéstető) installed systems. The two systems have been constructed to be identical. Detailed comparative statistical analysis may provide information about the polarization effect and the frequency dependent dumping of the surrounding andesite block. To achieve more accurate estimation of the attenuation characeristics, the surface observation instruments has been installed in the area of the Piszkéstető station which is an offsite observation station of the Konkoly Observatory, Research Centre for Astronomy and Earth Sciences, HAS. It lies about 5 km from the MGGL. In that scale the natural geomagnetic variation of the ULF-ELF frequency range (3-3000~Hz) can be considered to be homogenouos assuming similar induced contribution to the total magnetic field variation. Therefore the comparative analysis is reasonable in point of the two electromagnetic observing station.

\subsection{The data acquisition system}

\begin{figure}
\begin{centering}
\includegraphics[width=0.8\columnwidth]{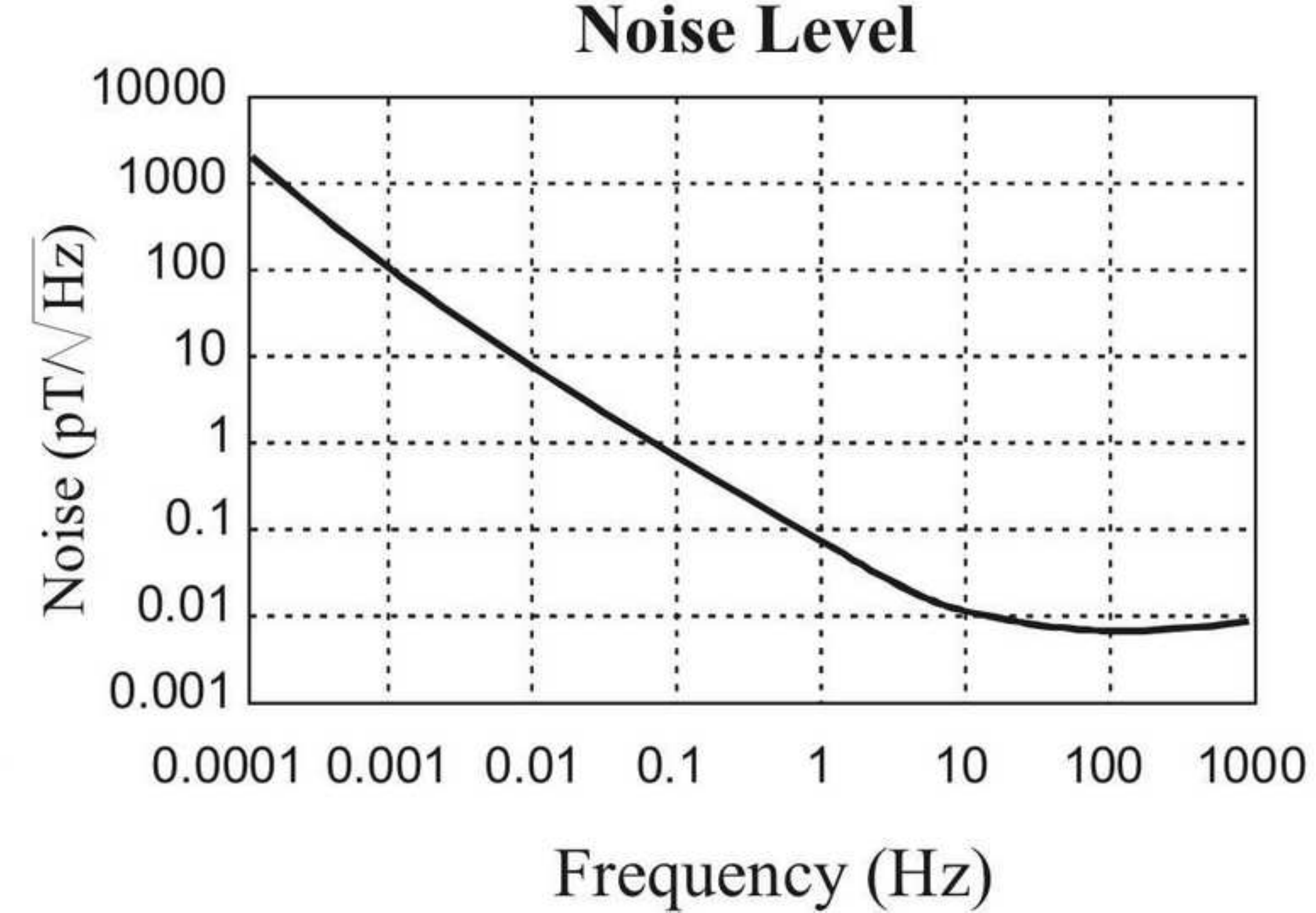}
\par\end{centering}
\protect\caption{Noise characteristics of Lemi-120 induction magnetometer (factory datasheet)}
\label{leminoise}
\end{figure}

The data acquisition system consists of a Raspberry Pi based data logger, two 16 bit A/D converters, an analogue filter and two induction coils as sensors of magnetic field variation positioned perpendicularly in a horizontal plane. The timing and synchronization is based on Network Time Protocol (NTP). As the main consequence of a few months test measurements shows dominant power line disturbances at 50~Hz and its odd harmonics, an attempt were made to attenuate the man-made noise and enhance the relative strength of natural origin components. 
In the course of the data acquisition magnetic variations of the upper ULF-ELF range was detected by means of Lemi-120 extremely low noise and wide frequency dynamic range induction magnetometers. The noise characteristics of Lemi-120 induction coils are plotted on Fig. \ref{leminoise}. The sensor noise remains below 0.1 pT Hz$^{-1/2}$ in the wider focus range of 1-100~Hz.

\begin{figure}
\begin{centering}
\includegraphics[width=0.8\columnwidth]{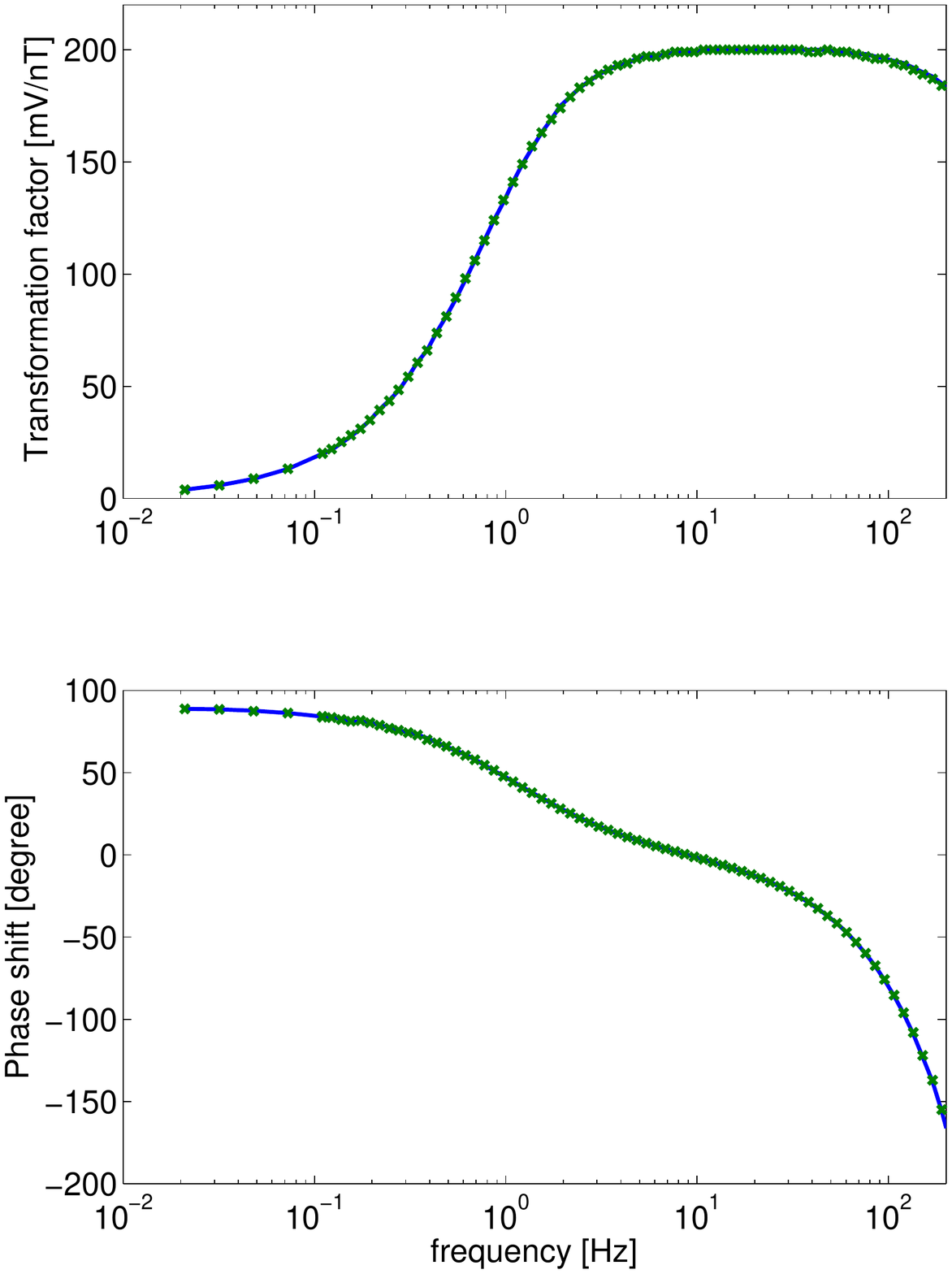}
\par\end{centering}
\protect\caption{Transfer function of the Lemi-120 serial number N649 and N650 induction coils installed in the MGGL. Green dots represents the transformation factor of the N650 sensor, while the blue curve is linear interpolation of the characteristic markers of N649 sensor.}
\label{trfvlemi}
\end{figure}

The amplitude transformation and phase shift of the two individual induction magnetometer are displayed on Fig. \ref{trfvlemi}. The green markers plotting the transfer function of the induction coil N650 fits  to the blue curve determined by induction coil N649 with less then 1\% tolerance.

The coils installed in the MGGL have been layed over a silica sand bedding to avoid coupling of microseismic noise through direct contact between the sensors and the concrete basement. The Lemi-120 induction magnetometers of the surface site have been buried 50~cm to avoid micro resonances of wind disturbances. The orientation of the sensors are exactly the same as in the MGGL. The hardware and the software system is just identical to the subsurface station.

\subsection{Results}
\begin{figure}
\begin{centering}
\includegraphics[width=0.8\columnwidth]{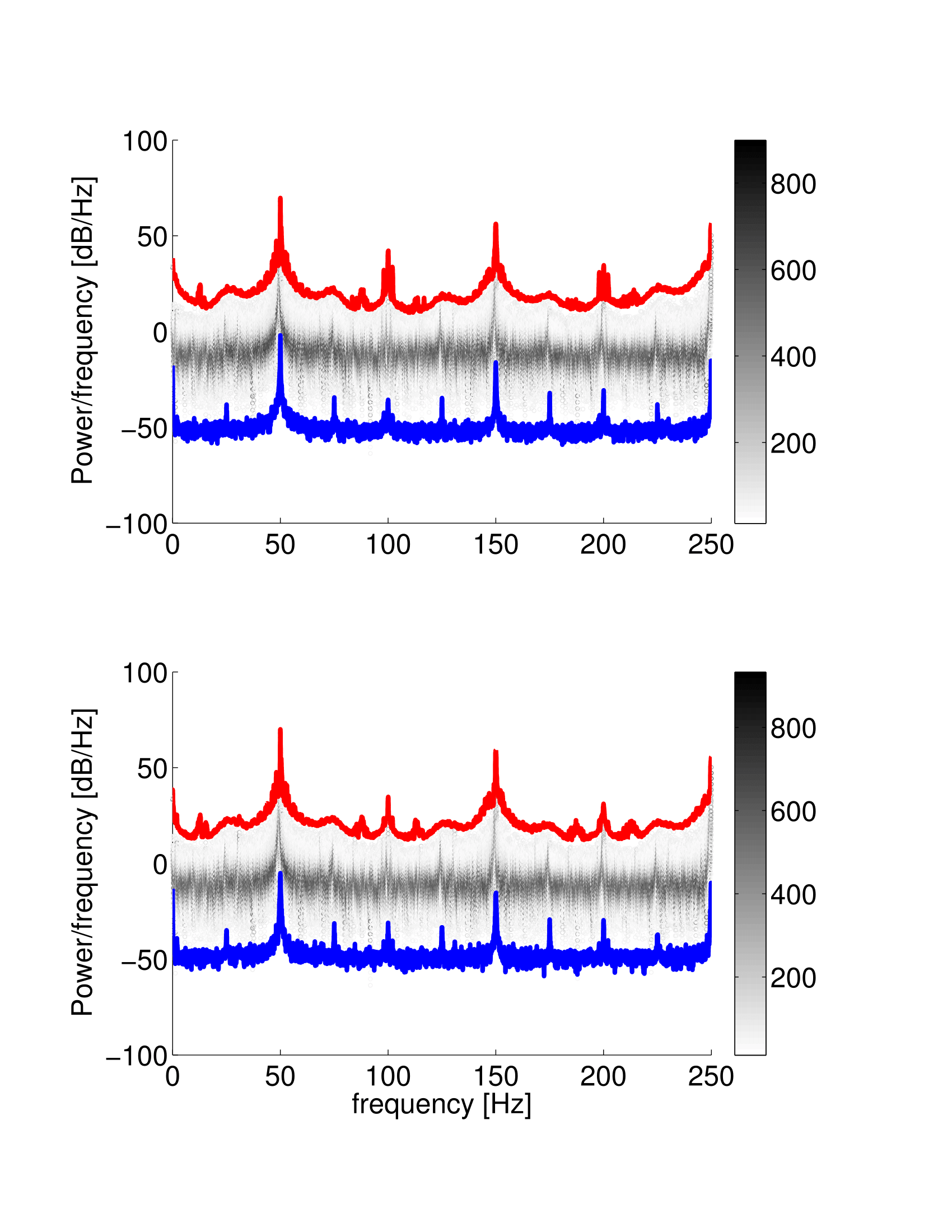}
\par\end{centering}
\protect\caption{Characteristic spectra of the north north-west oriented sensor (upper) and the east east-north oriented sensor (lower) installed in the MGGL. The red and the blue curves represent the highest and the lowest values occured in the studied representative time windows. The darkest gray shaded area represents the expected power spectral density.}
\label{spectra}
\end{figure}

\begin{figure}
\begin{centering}
\includegraphics[width=0.6\columnwidth]{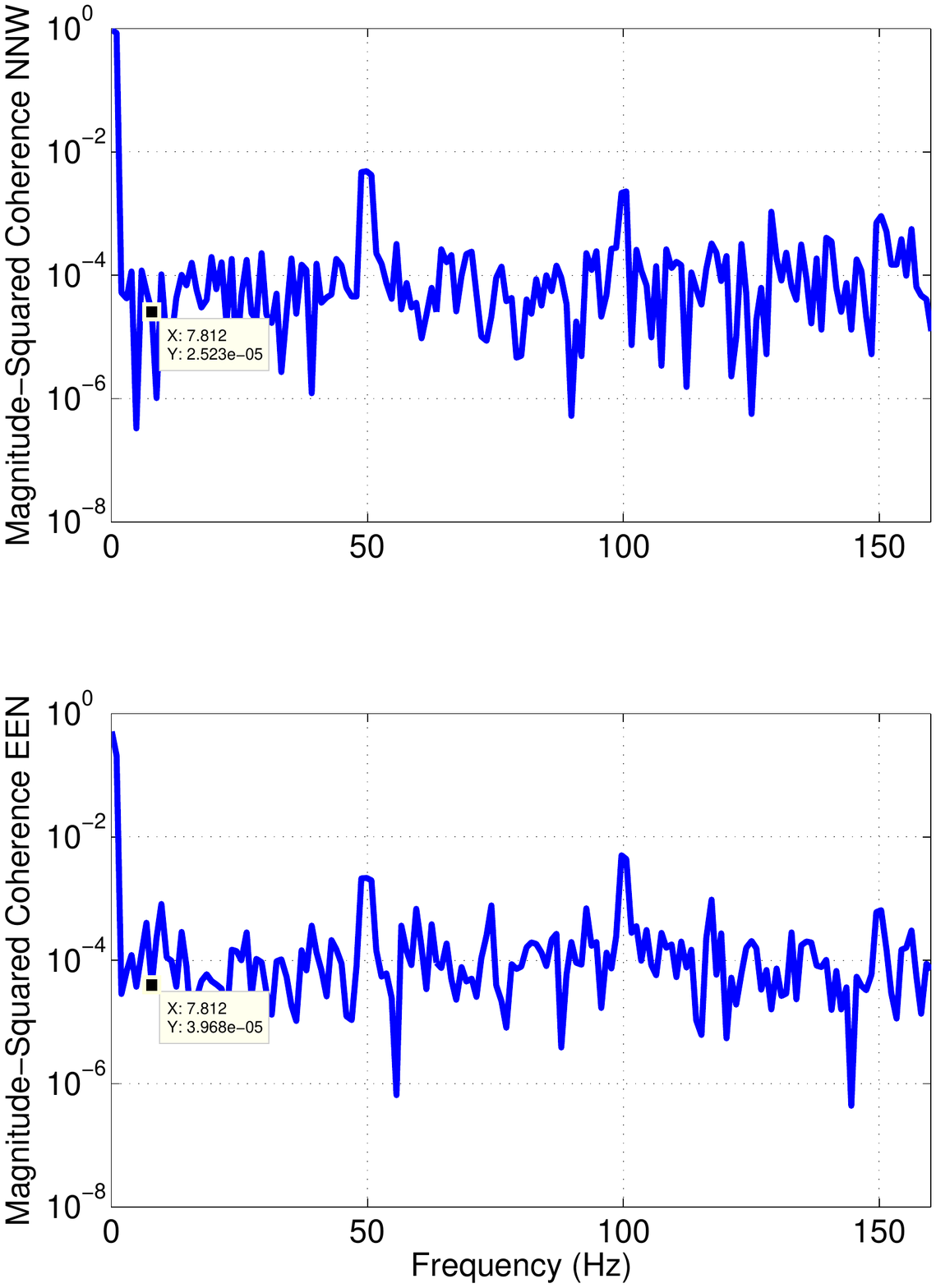}
\par\end{centering}
\protect\caption{Magnitude-square coherence of the time series' recorded at the MGGL and surface stations in the north north-west (upper) and the east east-north (lower) directed sensors,  based on a representative one hour long time window. Datatips show basic Schumann frequency.}
\label{MSC}
\end{figure}

The dynamic specra of the magnetic variations in the MGGL still demonstrates heavy load of power line disturbances and its harmonics, despite of the applied filter configuration, see surfaceplot on Fig. \ref{spectra}. The dynamic spectra of the environmental magnetic variation confirms that the power spectral density is continuously dominated by the 50~Hz and its upper harmonics, 100~Hz and 150~Hz. The spectral peaks of 16~Hz and 32~Hz is continuously present too. The noise loads are related to the power distribution of the MGGL and the suction pump which dredge up the water from the deep, and the lamp feeds of the mine respectively.
The lowest frequency mode of the Schumann resonance appears at about 7.83~Hz at low EM noise pollution observation stations. It is the strongest mode and slightly varies owing to the natural modulations of the resonator geometry. The magnitude-square coherence of the time series recorded at the MGGL and surface stations confirms that the natural origin spectral components cannot be identified by a 16 bit ADC besides the present conditions in either orientation (WSW-ENE and NNS-SSE), see Fig. \ref{MSC}. 
The location and the collected data seems not to be sufficient to let natural origin spectral components to be recognisable besides to man-made noise components. Relocation of the sensors, farther from the instruments mounted in the MGGL could significantly improve the signal-to-noise ratio and may facilitate comparative study of surface-subsurface dataset. Further improvement of the data quality will be achieved by the application of 24 bit ADCs, which is under developement. Implementation of the proposed modifications may result in improved quality of electromagnetic recordings to be feasible to extract natural origin spectral components.

\section{Measurements of the inhomogeneity of the rock-density at large-scale with cosmic muon tracking}
%Barnaf\"oldi G.G.$^1$, Hamar, G.$^1$, Ol\'ah, L.$^{1}$, Sur\'anyi G.$^{11}$,  Varga D.$^{1}$, Pázmándi, 

Cosmic-ray muons are created in the upper atmosphere and reach the surface of the Earth. Measurement of these highly penetrating particles opens the possibility to explore the large scale inner structure of rock formations above a measurement point by a detector placed underground. This technique was applied earlier for cave researches, cosmic background measurements for underground laboratories and for civil engineering~\cite{mt_nima:2012, mt_geo:2012, mt_ul:2013, mt_ecrs:2015, mt_npa:2016}. In the present study, this method was used to explore the inner structure of the M\'atra mountains above the MGGL, based on the idea that highly penetrating cosmic muons loose their energy at a known rate while passing through the medium. The measured muon flux carries information on the density-length (average density $\times$ thickness) of the medium, similarly as X-ray or computer tomography in case of medical or industrial usage with smaller size ($<$ meter) objects. The aim was to investigate the rock density homogeneity above the Gy\"ongy\"osoroszi mine by performing a high precision measurement of cosmic muon flux to cover the 2$\pi$ of the upper hemisphere. 

\subsection{The portable muon telescope}

To perform muon flux measurements in full 2$\pi$, a portable tracking system, called the Muontomograph, has been developed by the Wigner Research Centre for Physics \cite{mt_npa:2016}. The system, which uses Close Cathode Chambers~\cite{ccc:2011, ccc:2013}, is optimized for environmental applications~\cite{mt_nima:2012, mt_geo:2012, mt_ul:2013} with its weight of 15 kg, size of 37 $\times$ 33 $\times$ 27 cm$^3$, sensitive area of 25 cm by 25 cm, position resolution of 1.5 mm and angular resolution of 15 mrad. The Muontomograph is housed in a plexiglass box, which besides giving mechanical support, provides environmental isolation as well. 

\begin{figure}[!h]
	\centering
	\includegraphics[width=10cm]{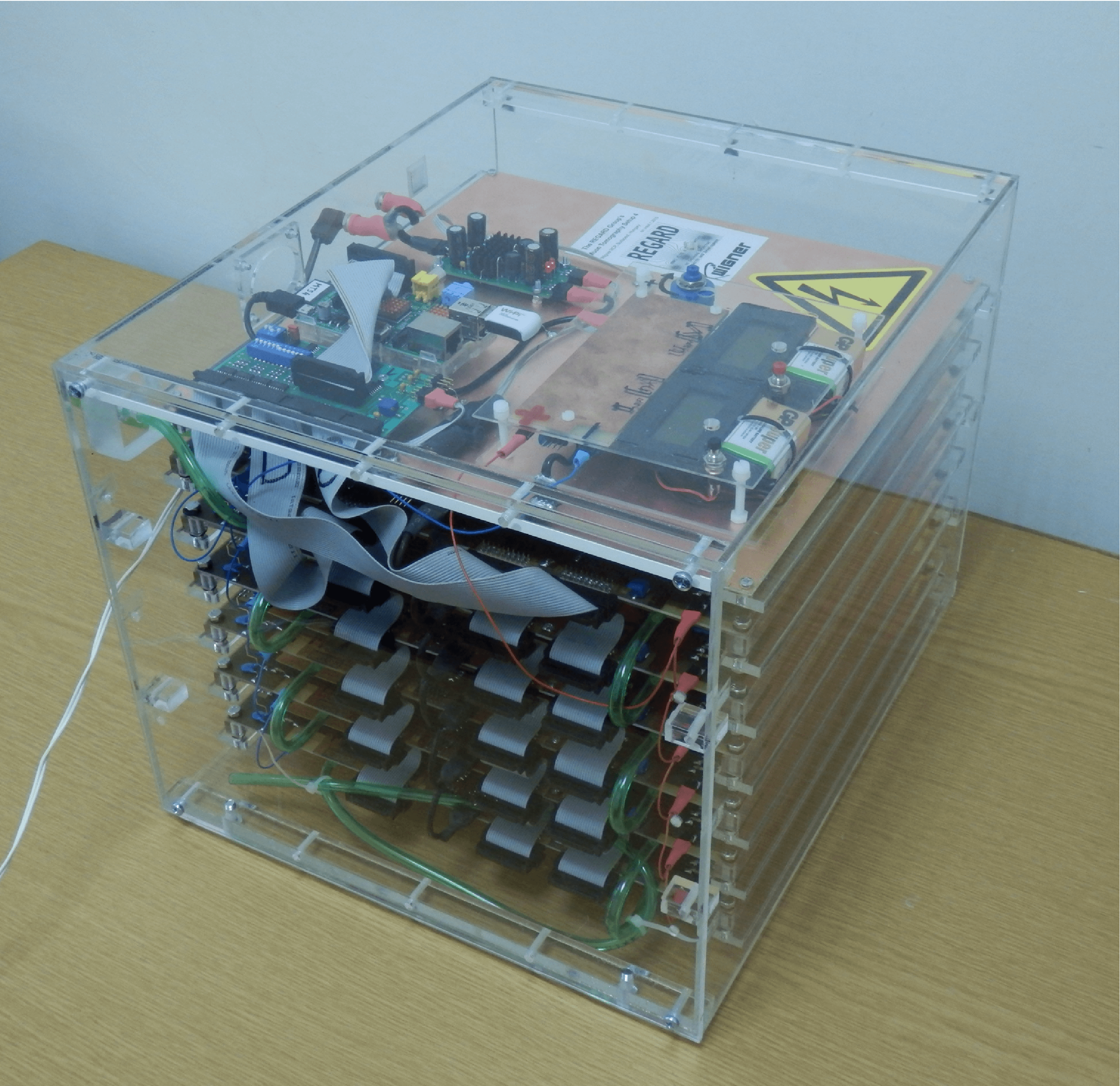}
	\caption{The portable muon telescope consisting of six Close Cathode Chamber tracking layers~\cite{mt_ecrs:2015}.}
	\label{fig:MTS4Structure} 
\end{figure}

The system requires continuous 0.5-2 l/h gas  flow during data taking, a standard, non-flammable mixture of Ar and CO$_2$ in 82:18 proportion. All functionalities and supply systems, including high/low voltage, triggering and data handling (DAQ) are integrated into one unit as on Fig.~\ref{fig:MTS4Structure} in Ref.~\cite{mt_ecrs:2015}. The measured track coordinates on the layers along with timing and triggering informations are recorded. The total power consumption of the complete detector system is about 6 W. With a standard, 10 l, high-pressure ($p_{max}<250$ bar) gas bottle, the system is capable of autonomous data taking for about a 2 months.

%The system requires continuous $0.5-2$ l/h gas flow during data taking, a standard, non-flammable mixture of Ar and CO$_2$ in 82:18 proportion. The Data Acquisition (DAQ) system is controlled by a single Raspberry Pi~\cite{raspberry}, and is placed above the upper tracking layer as presented on Fig.~\ref{fig:MTS4Structure}~\cite{mt_ecrs:2015}. All functionalities and supply systems, including high/low voltage, triggering and data handling are integrated into one unit. The measured track coordinates along with timing and triggering informations are written to a standard 16 GB Secure Digital (SD) card. The total power consumption of the complete detector system is about 6 W. With 10 l gas bottle the system is capable of autonomous data taking for about a 2 months.

%%%%%%%%%%%%%%%%%%%%%%%%%%%%%%%%
\subsection{Cosmic muon flux measurements in MGGL}

In order to determine the cosmic muon flux in the upper hemisphere, three measurements were performed with the Muontomograph in orthogonal directions at the same position in MGGL. The detector was located at the northwest corner of the laboratory room. First measurement was directed to the zenith and took 48.3 days. Second measurement was going on 41.9 days and directed to the east-northeast at ENE(66.5$^{\circ}$) and finally a third position with 34.5 days duration to the north-northwest NNW(335.5$^{\circ}$)~\footnote{Azimuth directions of the
detector-system readout reference were SSE(155.5$^{\circ}$) for Run-0-M2 and Run-0-M3, and for Run-0-M4 this was ENE(65.5$^{\circ}) $}. All measurements covered an angular region up to about 50$^{\circ}$ from the nominal direction, that is, providing a nice overlap between the settings. Details of the measurements are summarized in Tab.~\ref{tab:meas}. Data were downloaded from the detector via Ethernet connection during the data taking period in MGGL for online checking of data quality. Timing was provided by the NTP time protocol.

%%%%%%%%%%%%%%%%%%%%%%%%%
\begin{table}[!h]
\caption{\label{tab:meas} The summary table of the  RUN-0
data collection of MGGL in Gy\"ongy\"osoroszi mine: the
detector principal direction to the magnetic north and in
zenith, the duration of the measurements, number of events,
and the number of measured muon tracks.}
\begin{center}
\begin{tabular}{lcccccc}
\hline
 Runs & Direction & Zenith & Duration (days) & Events & Num.
of tracks\\
\hline
Run-0-M2& $ zenith $ & $ 0^{\circ} $ & 48.3 & 4.7 M     &
111,700         \\
Run-0-M3& $ ENE(65.5^{\circ}) $ & $ 90^{\circ} $ & 41.9 & 2.4
M & 18,124      \\
Run-0-M4& $ NNW(335.5^{\circ}) $ & $ 90^{\circ} $ & 34.5 &
3.1 M & 12,356  \\
\hline
\hline
\end{tabular}
\end{center}
\end{table}
%%%%%%%%%%%%%%%%%%%%%%%%

\subsection{Results of the rock inhomogeneity measurements}

\begin{figure}[!h]
	\centering
	\includegraphics[width=10cm]{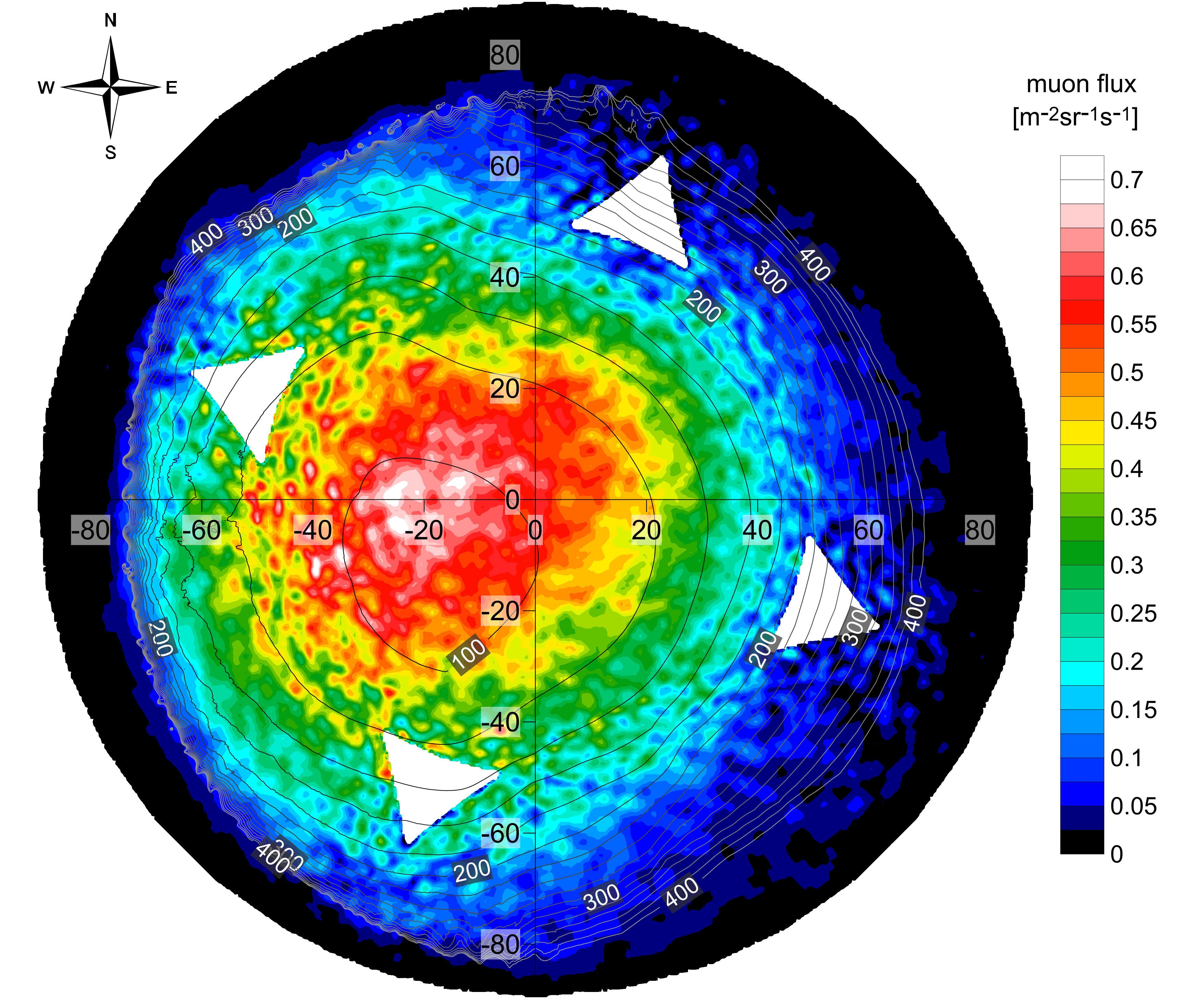}
	\caption{The cosmic muon flux map measured in MGGL is plotted as a function of azimuth and zenith angles from the detector position. Color-scale contours show the muon flux, dark contour lines show the detector-to-surface distance in meters. White arrows show the direction of the detector box.}
	\label{fig:MTLokpolTopo} 
\end{figure}

After analysing the measured data the muon flux was calculated based on the reconstructed tracks. The three measured and partially overlapping flux maps were merged together with the proper geometry and normalization~\cite{mt_npa:2016}. The obtained flux map is shown in Fig.~\ref{fig:MTLokpolTopo}, where the muon flux is plotted by color-scale contours as a function of azimuth and zenith angles. The three data settings, providing cosmic muon (track) rate of 0.005-0.02 Hz was sufficient to provide flux measurement with $5-50 \%$ statistical errors (depending on zenith angle). The measured flux has a maximum value of 0.7 m\textsuperscript{-2}sr\textsuperscript{-1}s\textsuperscript{-1} plotted with white color at $20^{\circ}$ zenith angle, exactly towards the West direction. The detector-to-surface distance is indicated with dark contour lines in the Figure~\ref{fig:MTLokpolTopo} with an estimated relative error of 5\% (5~m at the zenith).

One can observe that the surface distance and the muon flux reasonably match each other in Fig~\ref{fig:MTLokpolTopo}. With the present statistics of the muon flux measurement, the rock thickness, assuming homogeneous density, can be constrained to a precision of $10-25$~m. Within this relatively broad margin, the data do not show any large density inhomogeneities or cavities above the MGGL. Based on these promising results, the muon flux measurements will be continued to improve the rock density determination in the future.

\section{Summary and discussion}

The low frequency velocity of the andesitic rocks in the Gyöngyösoroszi mine is $c_l = 0.6455 \frac{km}{s}$, according to the data in section 2. The typical P and S wave speeds of andesitic rocks are $v_P = 5-6 \frac{km}{s}$ and $v_S = 2.8-3.4  \frac{km}{s}$ \cite{Zha16b}.  Therefore, the transfer function of a rock mass at low frequency are better approximated  by laboratory measurements of elastic moduli than the usual seismographic P and S wave velocity maps. A possible explanation is in the Appendix.

\begin{figure}[!h]
	\centering
	\includegraphics[width=15cm]{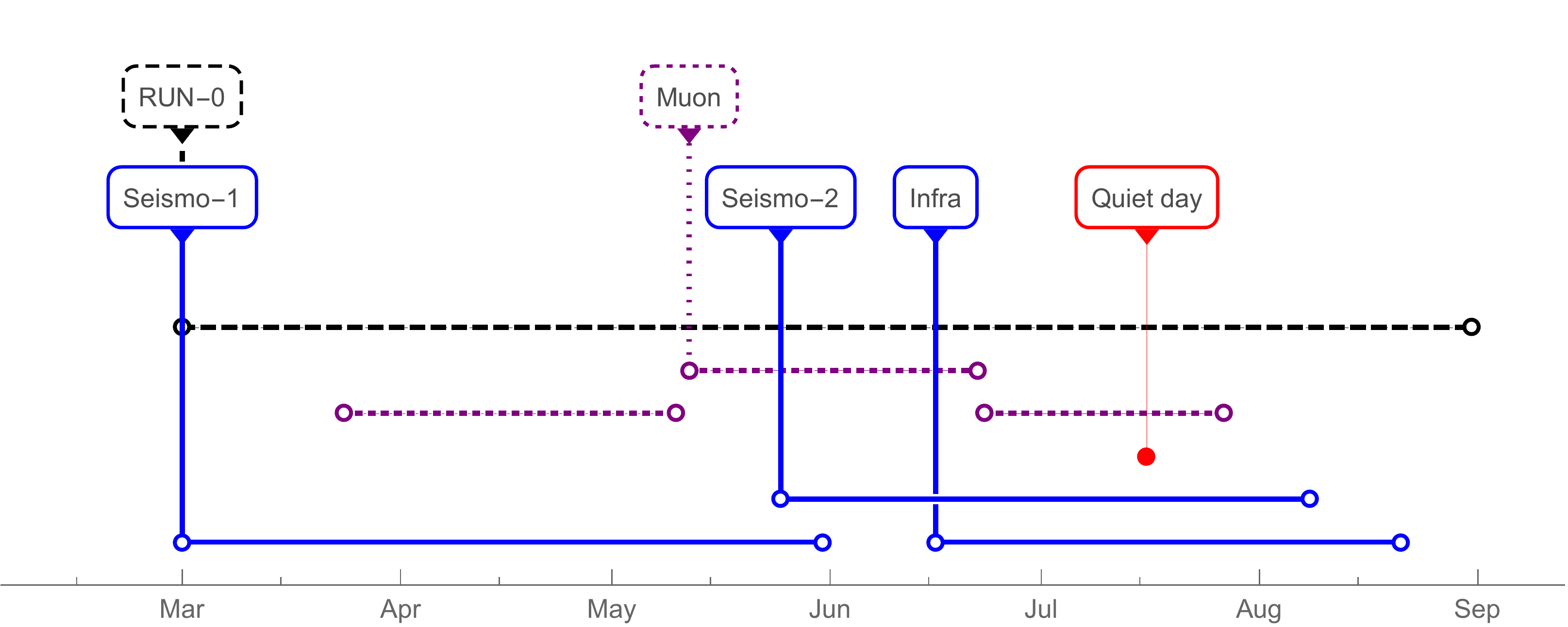}
	\caption{The timeline of the different instruments of MGGL during RUN-0 data collection.}
	\label{fig:timeline} 
\end{figure}

The timeline of the different instruments of MGGL during the 6 months of RUN-0 is shown on Fig \ref{fig:timeline}. 

During this test period the partially disjunct data acquisition periods restrict the possibilities of correlated and complex data analysis. In the next data acquisition period in 2017 (RUN-1) we expect reliable parallel data collection from all installed equipments, which would provide a detailed study and precise identification of different sources of low frequency noise. 

Seismic measurements have shown that the environment of MGGL is  more quiet than surface locations at the investigated frequency range (see Figures \ref{seismo2}, \ref{seismo3}). The noise spectra form MGGL were compared to the nearby seismological surface station at Piszkéstető. The preliminary site survey of Einstein Telescope assigned Mátra mountain range as one of the quiet locations in Europe \cite{ETdes11r,Bek13t}. During that study the seismic noise was measured for 3.5 days in two underground locations inside the Gyöngyösoroszi mine. The position of the less deep measurement was at 1435~m from the entrance, deeper in the mountain than the MGGL by 150~m. A qualitative comparison of the data of this previous registration period to our data with Guralp seismometer shows that the noise level of our recent long term measurements is slightly higher in the specific frequency range. However, a similar qualitative comparison to the measurements of the special seismometer of the Warsaw University indicates less difference. It is remarkable, that the noise level of the specially organized quiet day is close to the noise level of several other "normal" days, as one can see on Figure \ref{figwaw}. 

The seismic noise rms level from displacement spectra averaged in time and integrated for frequencies above 2~Hz was accepted as site noise characterisation number in \cite{BekEta15a}. During the test RUN-0 data collection at MGGL the rms level of the Guralp seismometer data in the quiet periods is similar that the rms level reported in the ET preliminary survey  in the less deep measurement location inside the Gyöngyösoroszi mine. One of these periods was e.g. the specially organized quiet day, 16\textsuperscript{th} July 2016, when recultivation activities and water pumping were ceased. Further investigations for the identification of natural and human activity related noise sources, like in \cite{NagEta14a}, are under progress.

The relatively high infrasound noise level (see Fig. \ref{ASD}), when contrasted to the low seismic noise, indicates that the human activities inside the mine did not filter into and do not propagate in the calm rock mass environment at the investigated frequencies.

%The possible reasons may be the periodic land reclamation activities and continuous water pumping near to the laboratory during the three month registration period. Easter in 2010!

The comprehensive geophysical study of the site involves detailed electromagnetic investigation of the local environment, too. Our preliminary noise measurements indicate a high-level electromagnetic noise that prevents the reliable determination of natural ones without increasing the effectiveness of filtering. We also plan a magnetotelluric exploration of the site to recover the spatial distribution of the electric conductivity in the close vicinity of MGGL. 

Based on the present data statistics, our underground muon flux measurement in MGGL indicates homogeneous density above the measurement location up to the precision of 10-15~m depending on the zenith angle.

\section{Acknowledgement}

The contribution and support of Nitrokemia Zrt. in particular Á. Váradi and V. Rofrits  is acknowledged. We also thank the construction work of Geofaber Zrt, and the remarks of Noa Mitsui. We thank our referees the detailed criticism, that  considerably improved our paper. 

This research was supported by MTA EUHUNKP grant. Authors PV and RK thank the  Hungarian OTKA and NKFIA grants K104260, K116197, K120660, NK106119, and K116375 for funding. TB, MC, MS, DR and TS were supported by the NCN Grant UMO-2013/01/ASPERA/ST9/00001. Muon flux measurements have been supported by the Momentum ("Lend\"ulet") grant of the Hungarian Academy of Sciences under contract LP2013-60. Authors GGB and MV also thanks the J\'anos Bolyai research Scholarship of the Hungarian Academy of Sciences and for the NewCompstar COST Action MP1304.

\section{Appendix: Rock mechanics and rheology}
%Ván Péter and Kovács Róbert
%Wigner FK and BME

{For the site selection of Einstein telescope the rock mass properties are interesting because the rock mass damps the external noises and also may be a source of noises. In both cases to understand the damping properties of various rock types at the relevant scale and relevant frequency range one needs to consider realistic material damping theories. Moreover, the knowledge of low frequency dispersion relation is the key of filtering the gravitational gradient noise. In this Appendix we shortly outline a simple model of acoustic  damping in rocks based on \cite{AssEta15a}. }

The wave damping of rocks is considered by a general exponential attenuation factor in seismology. That method focuses on amplitude and phase changes, without considering particular material models beyond elasticity \cite{AkiRic02b}. However, time dependent properties  of rock and also rock mass require more sophisticated models, and taking into account  material dependent dissipation both on large scales \cite{And89b,Sch06b} and in laboratory size samples \cite{Mats08a,LinEta10p,KovEta12a1}. Considering the difficulty and complexity of particular dissipation mechanisms in heterogeneous rock material, the construction of a detailed mesoscopic description is complicated, a universal mechanism independent model is preferable. That kind of theory is provided by local non-equilibrium thermodynamics with a single tensorial internal variable. This is the simplest consistent rheological constitutive relation, where universality is ensured by general thermodynamic principles \cite{Van13p2,AssEta15a}. Moreover, its suitability for time dependent rock characterisation is verified by the anelastic strain recovery (ASR) methodology, the available most sensitive in situ rock stress testing method \cite{Mats08a,LinEta10p}.

\subsection{Thermodynamic rheology}

Microstructural effects on mechanical properties of elastic materials is a traditional subject of  thermodynamic rheology \cite{Bio54a,Klu62a1}. In this case a single, local, second order and symmetric tensorial  internal variable should be introduced, which evolution couples to mechanical dissipation. Here a thermodynamical framework enforces a particular combination of inertia, relaxation and creep for dissipation for both the spherical and deviatoric components of the stress-strain relation. It is also remarkable that this complex material model arises with the simplest possible thermodynamic considerations, with a single internal variable without weak nonlocality. The theory leads to a fundamental thermodynamic rheological model, the  {\em Kluitenberg\,--\,Verh\'as body} \cite{AssEta15a}.

The elimination of the internal variable leads to the following expression between stress, $\sigma$ and strain, $\epsilon$:
\begin{equation}
\tau\dot \dst + \dst = E_2 \ddot \str  + E_1 \dot \str  + E \str,
\label{sKLb}\end{equation}
The coefficients have clear physical interpretation: $\tau$ is the stress relaxation time, $E$ is the modulus of elasticity, $E_1$ is a modulus of viscosity and $E_2$ represents material inertial effects. Every coefficient is nonnegative, moreover 
\begin{equation}
E_1 -E\tau > 0,
\label{rescon}\end{equation}
due to the nonnegative entropy production and concave entropy. The special cases must be analysed considering the original, thermodynamic coefficients, because the empirical ones are not independent. E.g. from the requirement of $\tau=0$ follows that $E_2=0$ and simple stress relaxation is forbidden.

The relevance of the Kluitenberg\,--\,Verhás material model for rocks is confirmed by the ASR method. There rheological properties of borehole rock samples are measured and analysed to determine in situ underground stress conditions. The application of the proper rheological model is the key of the quantitative evaluation. The basis of the  experimental analysis in the most reliable methodology is the complete Kluitenberg\,--\,Verhás body both for deviatoric and spherical parts of the stress--strain relation \cite{MatTak93a,Mats08a,LinEta10p}.

\subsection{Dispersion relations of the Kluitenberg\,--\,Verhás medium}

The Kluitenberg\,--\,Verhás body is best understood as a time-hierarchical two level Kelvin\,--\,Voigt system, as it is seen from the following rearrangement of \re{sKLb}.
\begin{equation}
\tau \frac{\d}{\d t}(\dst - (I_2+1) I_1  \dot \str - E \str) + (\dst - I_1  \dot \str - E \str) =0 ,
\label{sKLb_hie}\end{equation}

Where $I_1 = E_1-\tau E$ is the {\em index of damping} and $I_2= E_2-\tau I_1$ is the {\em index of inertia} \cite{AssEta15a}. This form shows well, that $\tau =0$ is not necessary for linear viscoelastic behaviour. If $I_2=0$, then the same linear viscolastic Kelvin-Voigt body appears in both in the first and the second terms of equation \re{sKLb_hie}, therefore the stress/strain relation is classical linear viscoelastic creep. Otherwise  the tendency to equilibrium can be damped oscillation or exponential relaxation depending on the sign of $I_2$ \cite{AssAta08a,Ful08a,Ful09a,AssEta15a}. 

It is straightforward to calculate the dispersion relation of the coupled thermo-rhelogical-elastic continuum in one spatial dimension, considering the basic balances, together with the Kluitenberg-Verhás material model. One obtains the following formula for the square of the phase velocity:
\begin{equation}
c^2 = \frac{E_1 \omega + \mathbf{i}(\omega^2 E_2 - E)}{\rho (\tau \omega - \mathbf{i})} = 
\frac{1}{\rho} \frac{(E+ \omega^2(\tau^2 E -I_2)) - \mathbf{i}\,\,\omega ( I_1+ \omega^2\tau(\tau I_1+I_2))}{\omega^2\tau^2 + 1}. 
\label{dre}\end{equation}

Both the dispersion relation, \re{dre}, and the original constitutive relation,  \re{sKLb_hie}, show two propagation speeds in the medium. The propagation speed at low frequencies, $c_l = \sqrt{\frac{E}{\rho}}$,  is different from the propagation speed at high frequencies, $c_h = \sqrt{\frac{E}{\rho}-\frac{ I_2}{\rho\tau^2}}$. In these two marginal cases the waves are not damped and in between there may exist a particular frequency where the dissipation is minimal: $\omega_{dmin} \approx \sqrt{\frac{I_1}{\tau E_2}}$. Then the  propagation speed may be far from its zero frequency limit, $c_l$. The effect is similar to the  "window condition" of low temperature heat conduction, where the optimal experimental heat pulse length is determined from the minimal dissipation condition \cite{MulRug98b}. A more detailed analysis of this dispersion relation can be found in Refs. \cite{Ver85b,Ver97b}.

\bibliographystyle{unsrt}
%\bibliography{fracmech,VanP,termo,stmech}
%\end{document}

\end{document}